\documentclass[aps,pre,noeprint,superscriptaddress,amssymb]{revtex4-2} 
\usepackage{amsmath,graphicx,listings}
\usepackage{url,hyperref}
\usepackage[usenames,dvipsnames]{xcolor}
\hypersetup{colorlinks=true, linkcolor=BrickRed, urlcolor=blue!50!black, citecolor=blue!50!black}
\usepackage[capitalize]{cleveref}
\usepackage{cleveref}
\usepackage{siunitx}
\AtBeginDocument{}
\crefrangelabelformat{equation}{(#3#1#4)$-$(#5#2#6)}

\begin{document}
\title{Phase Ordering in a few O(n) Symmetric Models: Slow Growth, Mpemba Effect and Experimental Relevance}
\author{Wasim Akram}
\affiliation{Theoretical Sciences Unit and School of Advanced Materials, Jawaharlal Nehru Centre for Advanced Scientific Research, Jakkur P.O., Bangalore 560064, India}
\author{Nalina Vadakkayil}
\affiliation{Theoretical Sciences Unit and School of Advanced Materials, Jawaharlal Nehru Centre for Advanced Scientific Research, Jakkur P.O., Bangalore 560064, India}
\author{Sohini Chatterjee}
\affiliation{Theoretical Sciences Unit and School of Advanced Materials, Jawaharlal Nehru Centre for Advanced Scientific Research, Jakkur P.O., Bangalore 560064, India}
\author{Subir K. Das}
\email{Email of corresponding author: das@jncasr.ac.in}
\affiliation{Theoretical Sciences Unit and School of Advanced Materials, Jawaharlal Nehru Centre for Advanced Scientific Research, Jakkur P.O., Bangalore 560064, India}
\date{\today}

\begin{abstract}
We study phase ordering dynamics in the three-dimensional nonconserved XY model, via Monte Carlo simulations, for quenches from paramagnetic phase to certain final temperatures $T_f$ within the ferromagnetic region of the phase diagram. The growth in the system occurs via annihilation of vortex and anti-vortex pairs, cores of which, in the three dimensional system geometry, join from different planes, on which the spins lie, to form line defects. In the long-time limit, the associated characteristic length scale, $\ell(t)$, appears to grow with time $(t)$ approximately as $t^{0.15}$, for $T_f=0$. The exponent is much smaller, like in the zero temperature intermediate time ordering in the three dimensional Ising model, than $1/2$, the expected value, that is realized for quenches to $T_f$ value that is sufficiently larger than zero. We carry out quenches from different starting temperatures, $T_s$, that lie above the critical temperature $T_c$. It is observed that the systems with higher $T_s$ approach the final equilibrium faster. This resembles the puzzling Mpemba effect. We present similar results also from the simulations of the two- and three- dimensional Ising model. In the case of the 2D Ising model, we show that the Mpemba effect is observed only if the starting magnetization is restricted to a value close to zero. In $d=3$, on the other hand, for both the models, the effect appears even if the initial configurations at a given $T_s$ are chosen from the full distribution of magnetization. Thus, our results are of much experimental relevance. 
\end{abstract}

\maketitle
\section{Introduction}\label{sec:intro}
When quenched to the same sub-freezing temperature, faster conversion of a hotter body of liquid water, into ice, than a colder one \cite{mpemba_cool,Tang2023Mpemba,Burridge2020Mpemba,Ghosh2025,Teza2026MpembaReview,Bechhoefer2021,Auerbach1995,langmuir_skd,Jeng2006}, is referred to as the Mpemba effect (ME) \cite{mpemba_cool}. In recent years similar effects have been observed in many other systems \cite{Oraz,Oraz_prx,avinash,chetrite2021metastable,Hartmut,vadakkayil,chatterjee,sohom_vicsek,granular_lasanta,granular_biswas,Chaddah2010,Carbon_nano2011,clathrate2016,Thermomajorization,Langevin_exp,QME_Joshi,QME_Amit,QME_Colin,QME_Zhang,QME_Ares,QME_Moroder,QME_Strachan,QME_XY}. Some specific examples are spin glasses \cite{baity}, antiferromagnets \cite{Oraz,Oraz_prx} and colloidal systems \cite{avinash,chetrite2021metastable,Hartmut}. 
In each of these, metastability is a key requirement for the exhibition of the effect. Furthermore, in the case of water \cite{Jin2015,Lasanta_water} importance of metastability was clearly demonstrated in a recent study \cite{Ghosh2025}. However, recently it has been argued that metastability is not essential for the appearance of the ME \cite{biswas2023mpemba,vadakkayil,langmuir_skd}. A simple, but physically relevant, situation that belongs to this category is the para- to ferromagnetic transition in the Ising model \cite{vadakkayil,langmuir_skd}. From the study of this model, in space dimension $d=2$, it is stated that differences in critical fluctuations in the initial states, at temperatures $T_s$, can lead to this puzzling effect \cite{vadakkayil,langmuir_skd}. 

It was argued that fluctuation-driven ME can also appear during magnetic transitions in the $q$-state Potts model \cite{chatterjee}, $q$ covering both first and second order transitions, fluid-to-solid transitions in Lennard-Jones (LJ) model \cite{Ghosh2025} and order-disorder transitions in velocity-aligning active matter systems \cite{sohom_vicsek}. 
However, in some of these studies, say, with the Ising and Potts models, the initial configurations were constrained by setting the order-parameter value to zero. While the appearance of the effect, despite the imposition of this restriction, is interesting, the results are of less significance from experimental point of view.
It should be noted that close to the critical point there can be significant fluctuations in order parameter. In experiments it is difficult to select systems with special restriction on instantaneous value of the magnetization.

In this work, we carry out study of kinetics of ordering in the nonconserved XY and the Ising models in $d=3$, via Monte Carlo (MC) \cite{landau_binder,newman1999,binder1992} simulations. In contrary to the previous study, with the 2D Ising or Potts models, here we consider full distributions of the order-parameter ($m$)  values at each $T_s$. Quite interestingly, the effect appears despite this unconstrained consideration. For analogous study in the 2D Ising model we show that the effect is nonexistent. We argue that this is due to differences in the critical fluctuations \cite{Fisher1967} with the variation of space dimension. We also present detailed simulation results, for both the models, by comparing the dynamics, in the asymptotic limit, for different final temperatures \cite{Puri2021}.

The remaining part of the paper is organized as stated below. In Sec.~\ref{sec:model} we describe the models and methods. The results are presented in Sec.~\ref{sec:results}. Finally, we conclude the paper in Sec.~\ref{sec:conclusion} with a brief outlook for future possibilities.
\section{Models and methods}\label{sec:model}

We write the general Hamiltonian \cite{landau_binder}, with nearest-neighbor interaction, as 
\begin{equation}\label{eq:hamiltonian}
 H=-J\sum\limits_{\langle ij\rangle} \vec{S}_i \cdot\vec{S}_j, \ 
\end{equation}
where $\vec{S}_{i(j)}$ is the spin at a lattice site $i(j)$ of a simple cubic lattice that we consider. For the XY model these spins are two component vectors with unit magnitude. For the Ising model, on the other hand, we have scalar spins that can take values $\pm1$.
 The interaction strength, $J$, is set to unity, in both the cases, that corresponds to transitions to ferromagnetic phases, from (disordered) paramagnetic phases. The critical temperatures for the Ising and the XY models have the values \cite{landau_binder,HASENBUSCH1990238,Gottlob,Hasenbusch2001} $T_c \simeq 4.51 J/k_B$  and $2.20 J/k_B$, respectively, $k_B$ being the Boltzmann constant. In the case of the Ising model, relevant defects are domain walls \cite{Bray,PuriWadhawan2009}, whereas for the XY model the ordering occurs via annihilation of vortices and antivortices \cite{Bray,Peter_Wills,chaikin1995}. As stated earlier we also present representative results for the 2D Ising model, from simulations on a square lattice. For this case we have \cite{landau_binder} $T_c\simeq 2.27 J/k_B$. Unless otherwise mentioned we study systems of linear dimensions $L=256$. We carry out MC simulations in which, for the XY model, a randomly chosen spin was given a trial rotation, $\Delta \theta$, within a range $[\SI{-0.1}{\radian},\SI{0.1}{\radian}]$ for the angle. These trial moves were accepted via the standard Metropolis algorithm \cite{landau_binder,Orkoulas2007}. The values of $\Delta \theta$ can be, in general, chosen from a larger or smaller range. For larger range, the acceptance rate of the trial move decreases, whereas, with a smaller rotation, the overall number of MC steps to reach the equilibrium state increases \cite{newman1999}. 
 Our choice of the above mentioned range for $\Delta \theta$ matches that of recent studies \cite{Puri2017,Puri2021}. The resulting dynamics does not preserve the value of the order-parameter over time \cite{Glauber1963}. For the Ising model a trial move is a combination of random picking of a spin and its flipping.

To quantify spatial ordering, we calculate the two point equal time correlation  function \cite{Bray,BP,Toyoki}, defined as 
\begin{equation}
    C(r,t)=\langle \vec{S}_i(t)\cdot \vec{S}_j(t)\rangle-\langle \vec{S}_i(t)\rangle\cdot \langle \vec{S}_j(t)\rangle,
\end{equation}
 $r$ being the scalar distance between sites $i$ and $j$. Typically, for a self-similar growth, the correlation function exhibits the scaling \cite{Bray,PuriWadhawan2009}:
 \begin{equation}\label{eq:eqn3}
     C(r,t)\equiv \tilde{C}(r/\ell),
 \end{equation}
where $\tilde{C}(x)$ is a master function that is independent of time.
We characterize the dynamics using the time dependent domain length $\ell$, which, for the XY model, is calculated from the decay of the correlation function to the value $0.1$. This, for the Ising model, has been calculated  from the first moment of the domain size distribution, defined as \cite{Majumdar,roughening_NV}
\begin{equation}\label{eq:domain_l}
     \ell(t)=\int P(\ell_d,t)\ell_d d\ell_d,
\end{equation}
where $\ell_d$ stands for the lengths of various domains which are estimated from the separations between consecutive interfaces along different Cartesian directions. We also calculate potential energy $E$ for the quantification of the evolutions.

 All presented quantitative results for dynamics in the XY model are averaged over $1450$ and $500$ independent initial configurations, respectively, for $T_f=0$ and $T_f=0.5T_c$. In the case of Ising model the quantities are averaged over a minimum of $1000$ initial configurations. 
 Unless otherwise mentioned all our results are for $d=3$. Also, if not mentioned, the results are without the $m \simeq 0$ restriction on the initial configurations.

\begin{figure}[h!]
\centering
  \includegraphics*[width=0.48\textwidth]{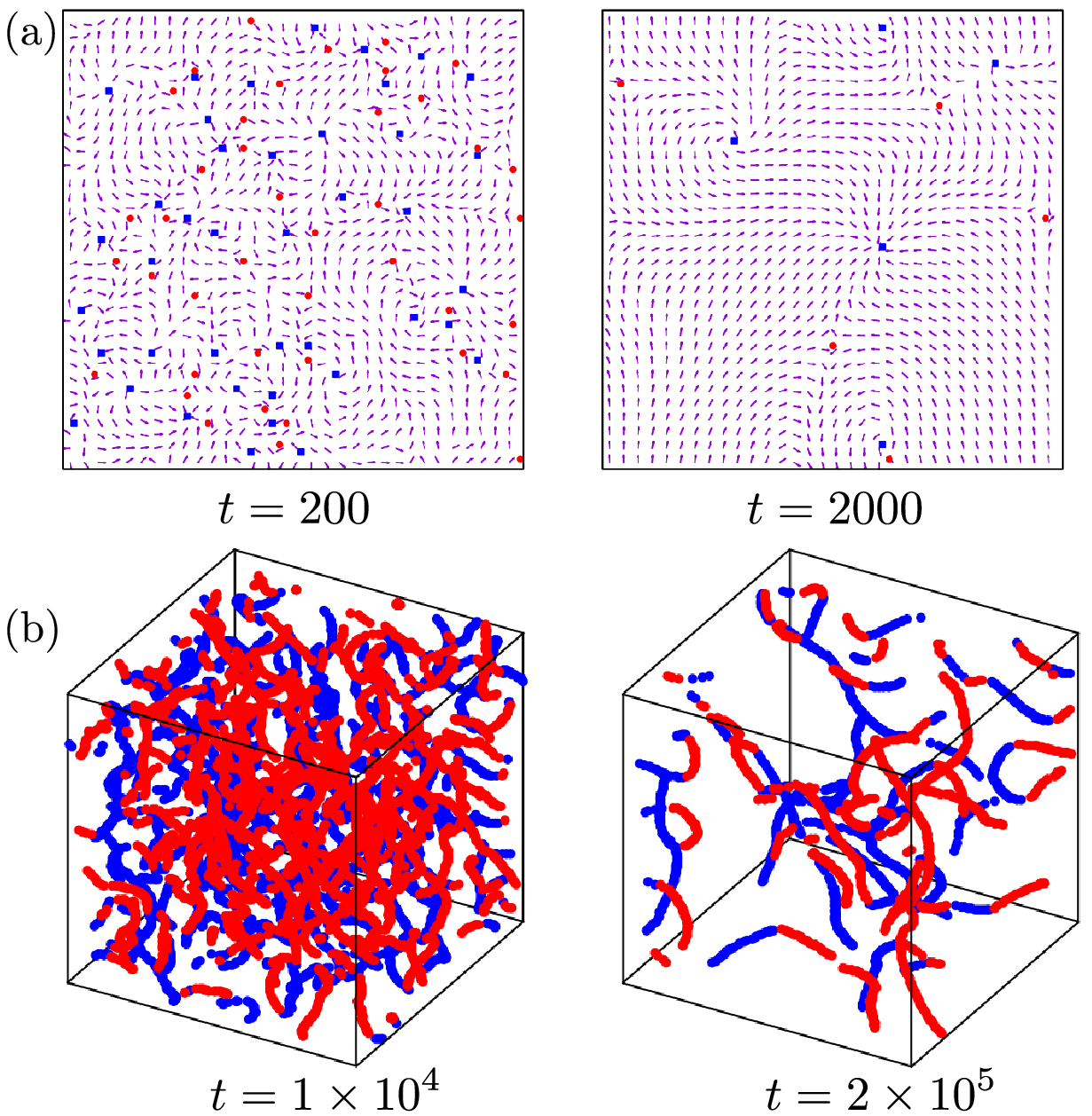}
 \caption{(a) 2D cross-sections showing spin configurations on a given plane of a few evolution snapshots for the XY model. These were obtained following a $T_f=0$ quench of an initial configuration with a random arrangement of spins, mimicking $T_s=\infty$. The arrows show spin orientations. The vortices and antivortices are shown in blue and red colors, respectively. (b) Similar to (a) but here we present 3D snapshots showing only the vortex and antivortex lines. Read related description in the text. In (a) and (b) the considered system sizes are $L=64$ and $L=256$, respectively. All further results are from simulations with $L=256$.}
  \label{fig:fig1}
\end{figure}
\section{Results}\label{sec:results}
\begin{figure}[h!]
    \centering
    \includegraphics[width=0.48\textwidth]{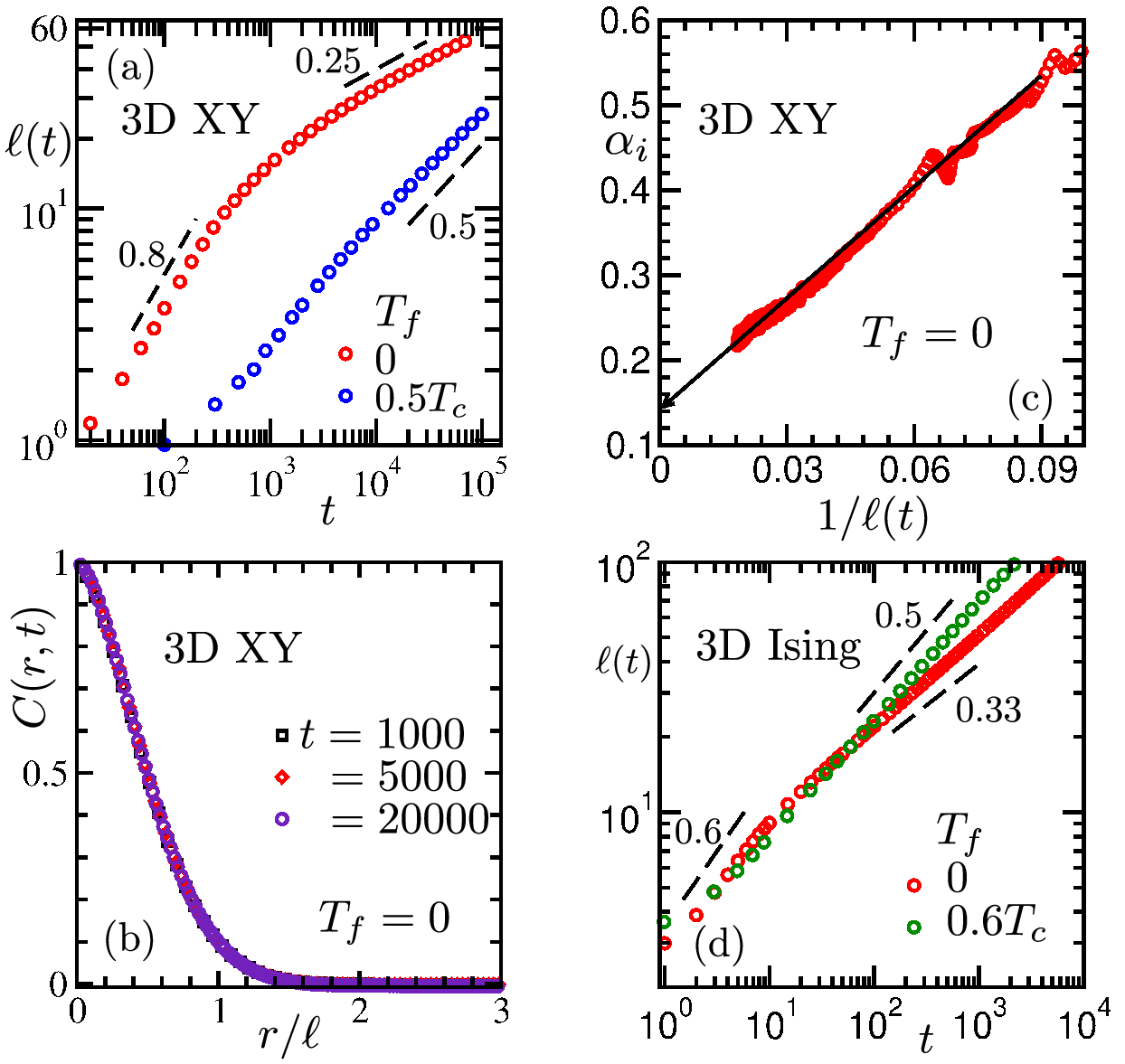}
    \caption{(a) Average domain lengths, $\ell(t)$, for the 3D XY model, obtained for quenches of initial configurations with random spin arrangements, imitating $T_s=\infty$, to $T_f=0$ and $T_f=0.5T_c$, are plotted versus time ($t$). (b) Scaling property of the correlation functions, for $T_f=0$, is demonstrated by using data from different times during evolutions according to the 3D XY model. Similar property holds for $T_f\ne 0$. In each of the cases the master function is described by an analytical form obtained by Bray, Puri and Toyoki \cite{BP,Toyoki}. (c) Instantaneous exponent, $\alpha_i$, corresponding to the growth with $T_f = 0$ in (a), is shown versus $1/\ell(t)$. The arrow-headed line there is a guide to the eye. (d) Domain lengths for the 3D Ising model, for $T_f=0$ and $T_f=0.6T_c$, are shown against time, following quenches from $T_s=\infty$. The data sets here are similar to  Ref. \cite{roughening_NV}. The dashed lines in (a) and (d) represent power-laws with the exponent values mentioned in the respective neighborhood. The errors in the data are smaller than symbol sizes.}
    \label{fig:fig2}
\end{figure}

In Fig.~\ref{fig:fig1} we show snapshots from the evolution of the 3D XY model following the quench of a perfectly homogeneous spin configuration, that corresponds to starting temperature $T_s=\infty$, to the final temperature $T_f=0$. In part (a) of this figure we present two-dimensional cross-sections, showing one of the planes on which the spins lie. The arrows there mark the orientations of individual spins and the symbols represent the locations of either vortices or antivortices. Location of such a point is determined by the condition \cite{SKD_pre2013}
\begin{equation}
    \oint \Delta\theta dl=2\pi p,
\end{equation}
 where $\Delta \theta$ is the phase difference between two adjacent spins on a (smallest) square plaquette and $p$ is an integer. For the present problem $p=\pm1$, corresponding to topological charges with different signs. As the defect density decreases, with time, via annihilation of vortex-antivortex pairs \cite{Goldenfeld_2d}, the (expected) enhancement of ordering of the spin directions is clearly visible in Fig \ref{fig:fig1}(a). Note that the defect cores from different parallel planes join to form lines \cite{Goldenfeld_3d}. In part (b) we present snapshots showing evolution in terms of these line defects. It should be noted here that the dimensionality of a defect \cite{Bray} in $d$ space dimensions with order-parameter symmetry $n$ is $d-n$. Clearly, the density of lines are decreasing with time, implying growth. While the snapshots in Fig.~\ref{fig:fig1}(a) correspond to $L=64$, the ones in Fig.~\ref{fig:fig1}(b) are for $L=256$. From here on all the results are for $L=256$.

In Fig.~\ref{fig:fig2}(a) we show the characteristic length, $\ell(t)$, versus $t$, related to the quench protocol and model in Fig.~\ref{fig:fig1}, viz., from $T_s=\infty$ to $T_f=0$ for the 3D XY model. There we have also included, for a comparison purpose, data for the combination $T_s=\infty$ and $T_f=0.5T_c$. Typically, for such spin systems, $\ell(t)$ grows as \cite{Bray,Tauber2017}
\begin{equation}
    \ell\sim t^\alpha,
\end{equation}
with \cite{Bray} $\alpha=1/2$, when the dynamics is nonconserved \cite{Bray_Rutenberg,Blundell_Bray,Puri2017,Puri2021}, like in the present case. 
This growth law is seen to be obeyed fairly well \cite{Puri2021} by the data set for $T_f=0.5T_c$. However, the late time data for $T_f=0$ is clearly in significant deviation from this expectation! Before presenting further quantitative results on this slow growth, in Fig.~\ref{fig:fig2}(b) we demonstrate scaling \cite{Bray} in the structural property for the $T_f=0$ case. There we have plotted $C(r,t)$ as a function of $r/\ell$. Nice collapse of data from different times imply self-similar growth \cite{Bray,PuriWadhawan2009}, that we discussed above. See Eq.~(\ref{eq:eqn3}).

In Fig.~\ref{fig:fig2}(c) we show the instantaneous exponent \cite{Huse,Majumdar}, 
\begin{equation}
    \alpha_i=\dfrac{d\ln \ell}{d\ln t},
\end{equation}
as a function of $1/\ell$, for the growth data set with $T_f=0$. In the limit $1/\ell\to 0$, the convergence in such a plot is expected to provide the asymptotic growth exponent. In this case, it appears that $\alpha\simeq 0.15$! We have checked via simulations of systems of different sizes that such a low value of the exponent is not a consequence of finite-size effects. A slower than expected growth was also reported for the 3D Ising model \cite{zerotmp_redner,zerotmp_NV,roughening_NV,Janke_zero,Chakraborty2017,Redner2001} with $T_f=0$. Representative results from this case are shown in Fig.~\ref{fig:fig2}(d), for $T_f=0$ and $T_f=0.6T_c$, with $T_s=\infty$. For the $T_f=0$ case, clearly there exists an extended time regime for which the exponent is significantly smaller than $1/2$. At late enough time, however, the data exhibit a crossover to $\alpha=1/2$ behavior \cite{roughening_NV,Das2017,zerotmp_NV}. This is to be seen whether such a crossover picture is true even for the 3D XY model. However, with the available resources at present, such a study is beyond our capability. This is because of the requirement of simulations of much larger systems over significantly longer period.

\begin{figure}[h!]
    \centering
    \includegraphics[width=0.48\textwidth]{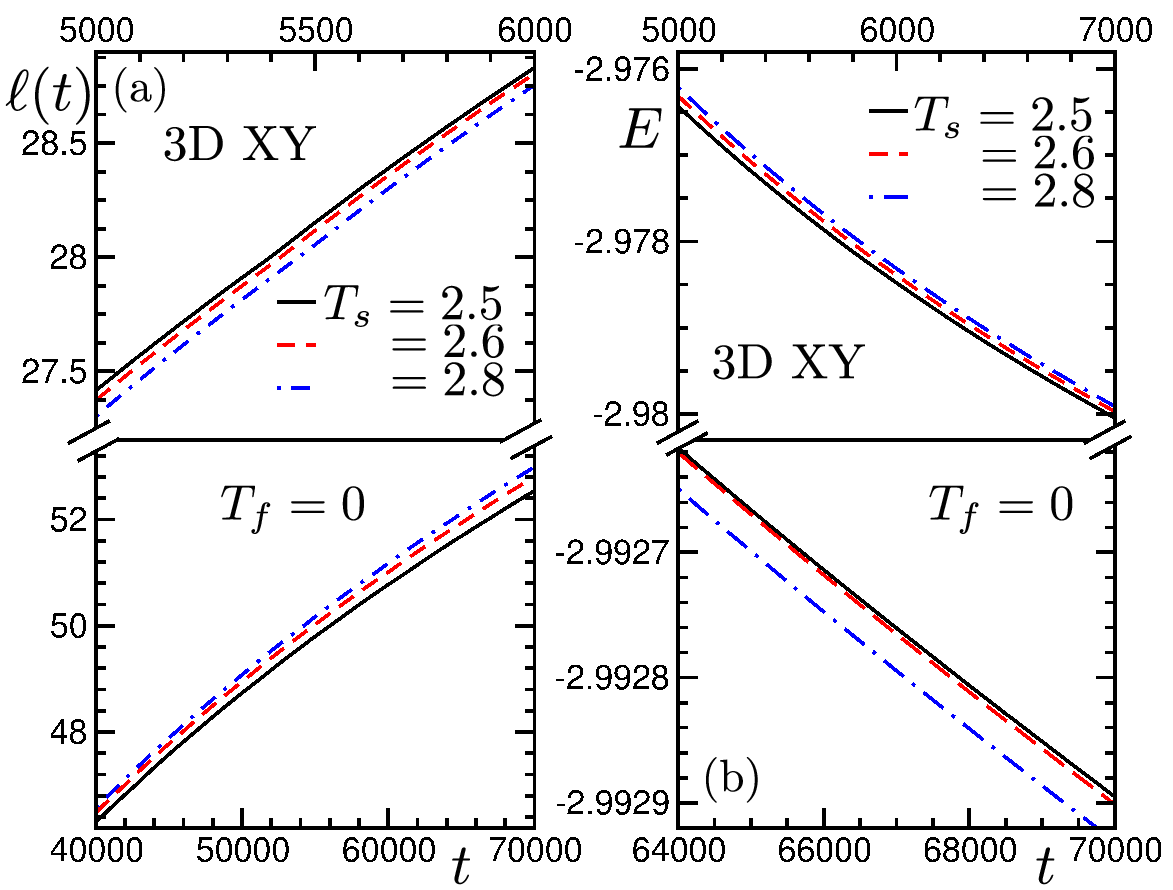}
    \caption{(a) Average domain lengths for the 3D XY model are plotted versus time for different starting temperatures $T_s$. (b) For the same $T_s$ values as in (a), we show the energy ($E$) versus $t$ plots for the 3D XY model. In both the cases, the upper sub-frames show the early time behavior and the lower frames show the late time behavior. These results are for quenches to $T_f=0$.}
    \label{fig:fig3}
\end{figure}

Next, we carry out quenches from several values of $T_s$, lying above $T_c$, to each of the above considered $T_f$, for both the models. For this purpose, starting from random initial configurations, following the same MC protocol, we equilibrate the systems at the desired values of $T_s$. Due to critical slowing down \cite{Hohenberg,landau_binder} this is a computationally demanding process when $T_s$ is close to $T_c$.
For this general quench protocol, first we present results for $T_f=0$. In Fig.~\ref{fig:fig3} we discuss the corresponding evolutions for the 3D XY model. In part (a) we show the $\ell$ vs $t$ plots. The frame has been divided into two sub-frames. In the upper sub-frame we show data from early period. The lengths appear larger for lower $T_s$. This sequence gets reversed in the lower sub-frame where the results are from the late times. This implies faster evolutions of systems having hotter start, resembling the puzzling Mpemba effect \cite{mpemba_cool}.
The same conclusion can be arrived at from the calculations of (potential) energy. These results are shown in Fig.~\ref{fig:fig3}(b). Expectedly, at early times systems with hotter starts have higher energies than the ones with colder starts. Again, the sequence gets reversed at later times, implying the presence of the ME. 

\begin{figure}[h!]
    \centering
    \includegraphics*[width=0.48\textwidth]{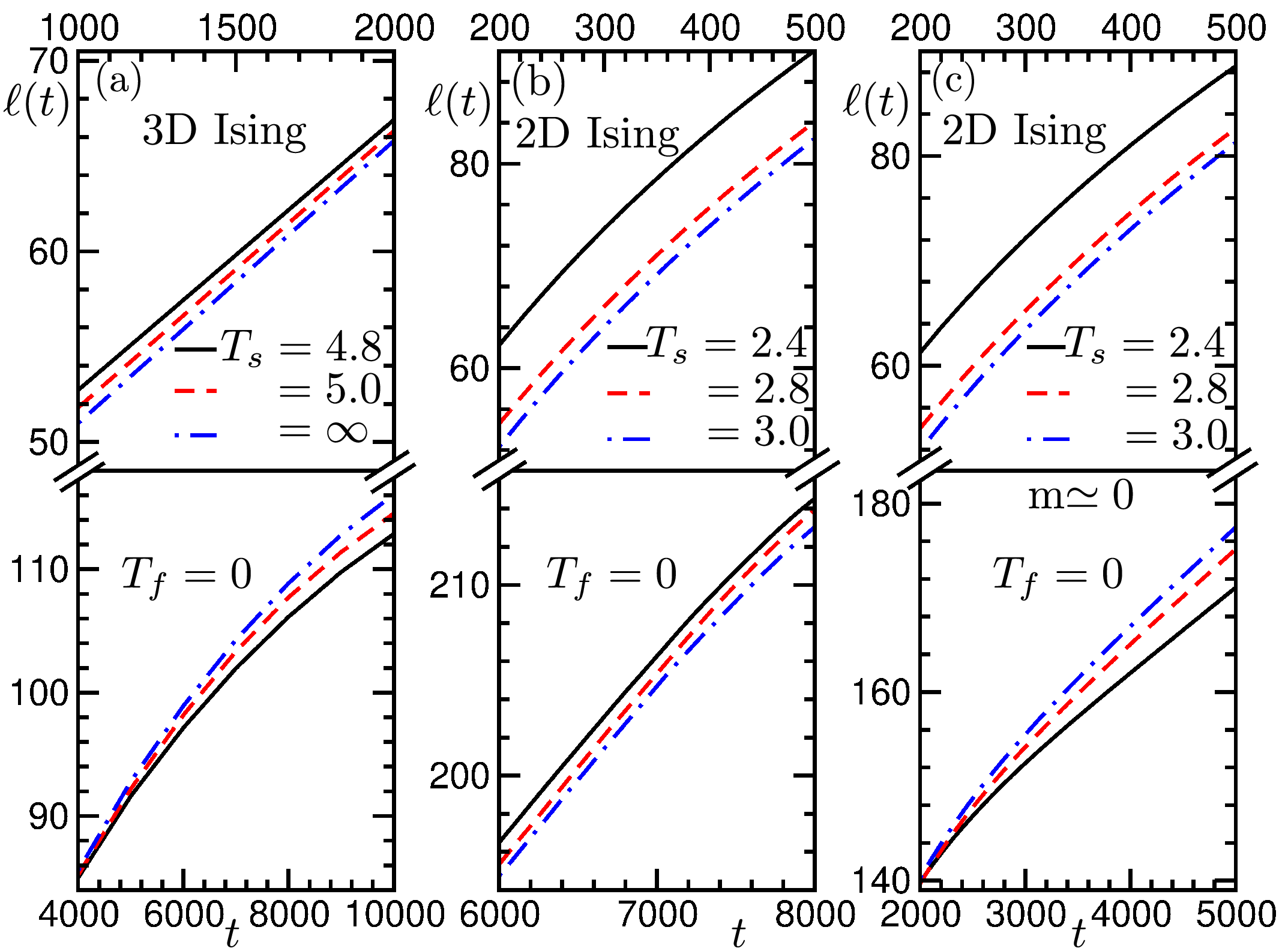}
    \caption{(a) Plots of $\ell(t)$, versus time, for the 3D Ising model, following quenches of initial configurations from a few different $T_s$ to $T_f=0$. (b) Same as (a) but here the results are for the 2D Ising model. (c) Same as (b) but in this case the magnetization in the initial configurations have been restricted to approximately zero. Recall that this restriction is used only for this particular case in our study.}
    \label{fig:fig4}
\end{figure}
Analogous results for the Ising model are presented in Fig.~\ref{fig:fig4}. For this model we include results from $d=2$ as well. For the sake of brevity, in this case we present results only for the domain growth. In Fig.~\ref{fig:fig4}(a) the data sets are from $d=3$. Clearly, crossings analogous to the 3D XY model are visible, implying the existence of the Mpemba effect. In Fig. \ref{fig:fig4}(b) the data are for the 2D Ising model. Mpemba effect cannot be detected till very late time by which growths of the systems are nearly complete.
However, in this case, when the initial magnetization is restricted to $m \simeq 0$ the Mpemba effect is observed ~\cite{vadakkayil,chatterjee}. For completeness, we show this scenario in Fig. \ref{fig:fig4}(c).

\begin{figure}[h!]
    \centering
    \includegraphics*[width=0.48\textwidth]{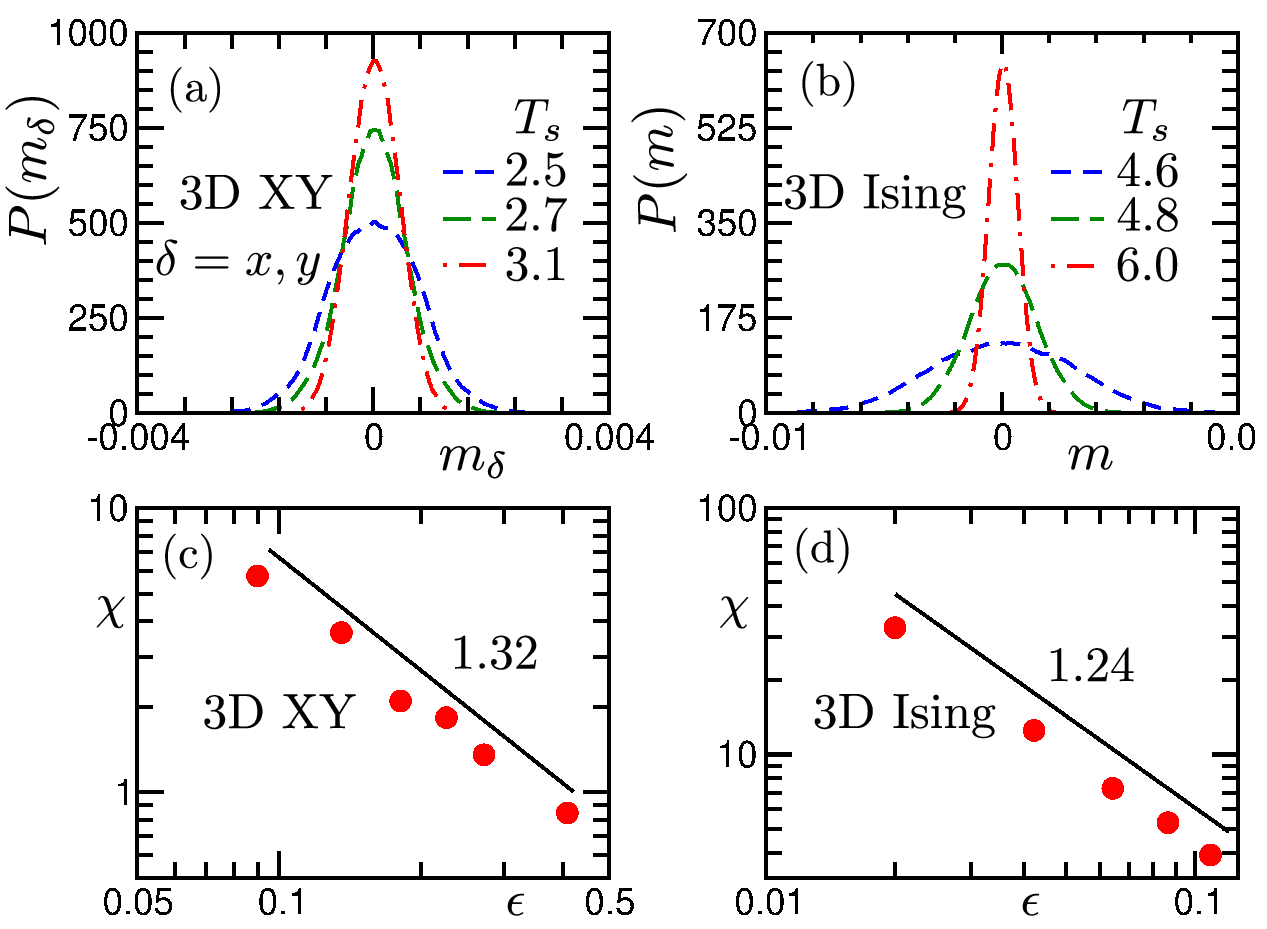}
    \caption{Distributions of order parameter at various $T_s$ for (a) the XY model and (b) the Ising model in $d=3$. In (a) the plots represent combined component-wise ($\delta= x,y$) distributions. The critical behavior of $\chi$ in the two cases are shown in (c) and (d). The solid lines there represent power-laws with exponent values mentioned inside the frames. The numbers for statistics for these static quantities are $2900$ and $80000$ for the XY and the Ising models, respectively.}
    \label{fig:fig5}
\end{figure}

It should be noted here that the magnetization of the systems at the initial states can fluctuate \cite{Fisher1967}. In the XY model the magnetization per spin is written as $\vec{m}=m_x\hat{i}+m_y\hat{j}$, where, $m_x = \frac{1}{N} \sum_{i=1}^N \cos \theta_i$ and $m_y = \frac{1}{N} \sum_{i=1}^N \sin \theta_i$, $\theta_i$ and $N~(=L^3)$ being the orientation of a spin and total number of spins, respectively.
Though well known, we demonstrate the fluctuation in Fig.~\ref{fig:fig5} for the few $T_s(>T_c)$. The distributions of each of the components of magnetization is equivalent to the other, viz., $P(m_x) \equiv P(m_y)$.
There, in part (a) we show distributions of magnetization from different $T_s$ for the 3D XY model. Similar results for the 3D Ising model are shown in part (b). In the latter case, $m=(1/N)\sum_{i=1}^N S_i$. As expected, the fluctuation clearly becomes wider for states that are closer to the critical point  \cite{Fisher1967,Stanley,Onuki2002}, implying critical divergence of susceptibility ($\chi$) that can be calculated from the standard deviations of the distributions. Here note that \cite{Fisher1967,landau_binder} $\chi\sim \epsilon^{-\gamma}$ and $\xi$, the spatial correlation length, $\sim\epsilon^{-\nu}$, with $\epsilon=|T-T_c|/T_c$.
For the sake of completeness we also have demonstrated the critical singularities of $\chi$ for these models in Fig.~\ref{fig:fig5}(c) and (d).
In computer simulations, typically one uses finite-size scaling method to obtain or confirm the expected values of the exponents. However, given that our system sizes are quite large, in the presented temperature range the size effects are quite weak.

For an experimentally relevant Mpemba effect, one should consider configurations from the full distributions of the magnetizations, irrespective of the value of $T_s$, even if for some configurations, with magnetizations far away from zero, corresponding to $T_s$ very close to $T_c$, the race to equilibrium at $T_f$ is unfairly advantageous. However, for this protocol the Mpemba effect appears nonexistent, or at the very best only weakly existent, in the $d=2$ case. 
This can be due to the fact that fluctuation of order parameter in $d=2$ is broader than that in $d=3$. Note that the critical exponent $\gamma$, for susceptibility \cite{Fisher1964,Kadanoff,Fisher1967,ZinnJustin2001}, in $d=2$, is $1.75$, whereas the value is $\simeq 1.24$ in $d=3$. 
Now we return to the 3D cases for which initial configurations were chosen from the full distributions of the order parameter, at each $T_s$, as stated earlier.

\begin{figure}[h!]
    \centering
    \includegraphics[width=0.48\textwidth]{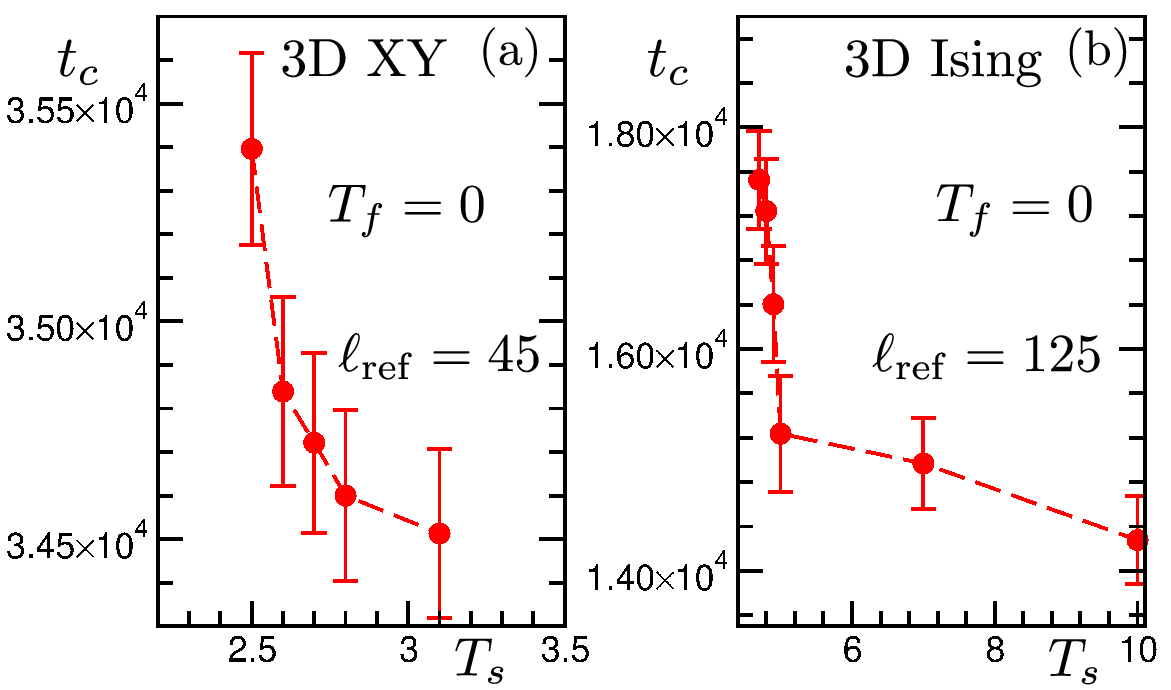}
    \caption{(a) The crossing times for the 3D XY model, with a given reference length, after crossings among the data sets for different $T_s$ have already occurred. (b) Same as (a), but for the 3D Ising model. These results are for $T_f=0$.}
    \label{fig:fig6}
\end{figure}

To confirm systematicity on the trends of evolutions from different $T_s$, in Fig. \ref{fig:fig6} we plot the corresponding crossing times, $t_c$, defined as the duration to reach a given reference domain length, $\ell_{\rm{ref}}$, after crossings among data for all different values of $T_s$ have occurred. In part (a), results from 3D XY model and in part (b), results from the 3D Ising model, both for quenches to $T_f = 0$, are presented. The errors there are calculated by using the Jackknife resampling method \cite{newman1999,efron_jackknife}. Irrespective of the model, overall decrease of mean value of $t_c$ with increase in $T_s$ can be appreciated, which is a consequence of the presence of ME in these systems. The robustness of the effect depends upon the separation between initial temperatures.

\begin{figure}[h!]
    \centering
    \includegraphics[width=0.48\textwidth]{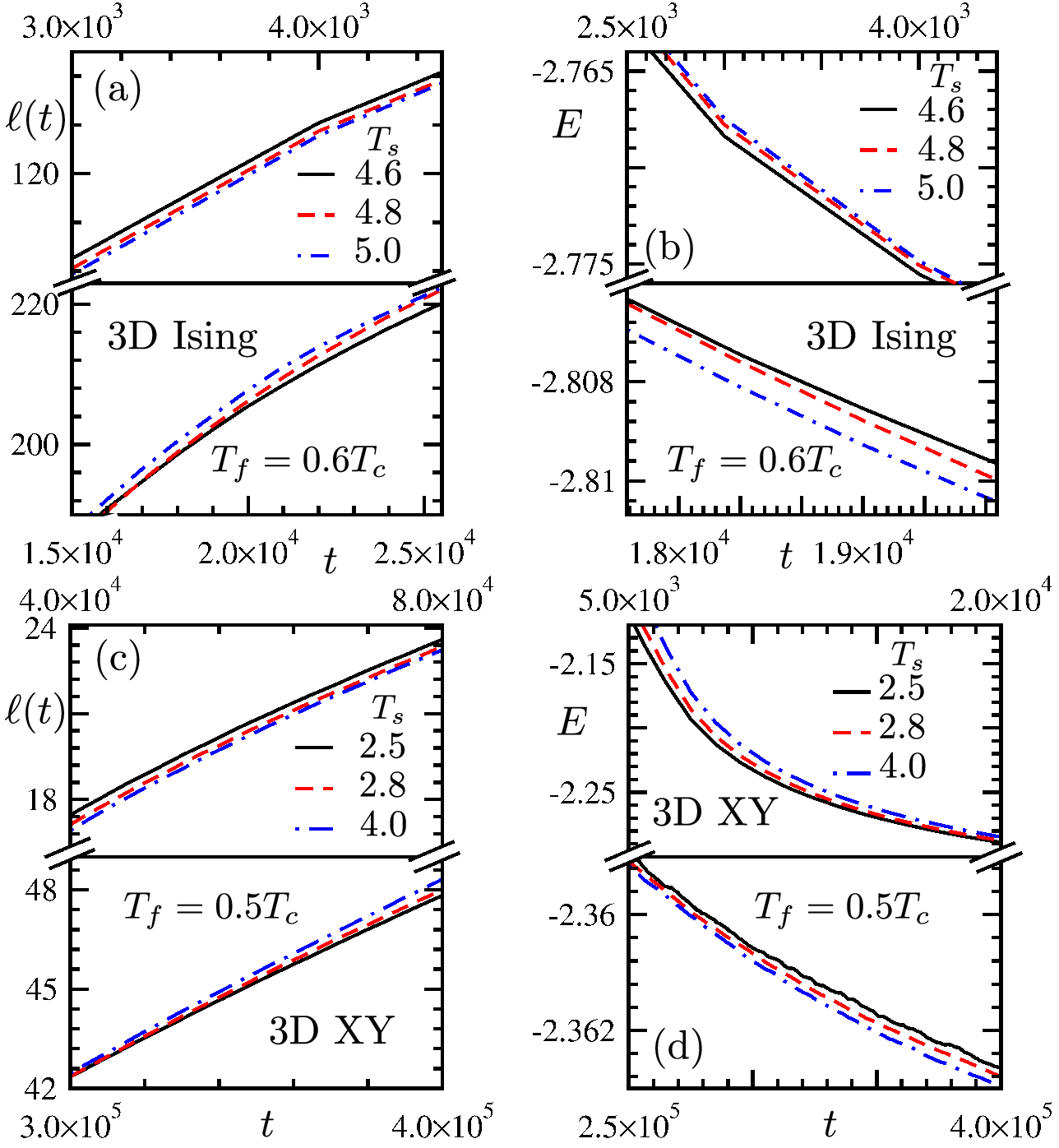}
    \caption{(a) Average domain lengths are plotted versus time, following quenches to $T_f=0.6T_c$, from different starting temperatures $T_s$, for the 3D Ising model. (b) Plots of $E$ versus time, corresponding to the conditions mentioned in (a). (c) Same as (a) but for the 3D XY model with $T_f=0.5T_c$. (d) $E$ versus $t$ plots for considerations in (c). }
    \label{fig:fig7}
\end{figure}

Here one may ask, if the Mpemba effect appears for evolutions following quenches to only very low temperatures. To answer this, in the following, for the models in $d=3$, we present results for $T_f \ne 0$. In part (a) and (b) of Fig.~\ref{fig:fig7} the results are, for $T_f=0.6T_c$, from the $3D$ Ising model. Trends in both domain lengths, see Fig.~ \ref{fig:fig7}(a), and potential energies, see Fig.~\ref{fig:fig7}(b), indicate the presence of the ME. In (c) and (d) of this figure we present analogous results for the 3D XY model with $T_f=0.5T_c$. The conclusion remains same here as well.

\section{Conclusion}\label{sec:conclusion}
We have presented results on kinetics of phase ordering \cite{Bray} for para- to ferromagnetic transitions in space dimension $d=3$. These are obtained via Monte Carlo simulations \cite{landau_binder} with the Ising and XY models. For this purpose we have quenched configurations from within paramagnetic phase to final temperatures, $T_f$, below the Curie temperature. For the XY model, it appears that the growth exponent in the asymptotic limit is approximately $0.15$, when $T_f=0$, much smaller than the expected value $1/2$. This picture has resemblance with an extended intermediate period growth scenario in the Ising model \cite{roughening_NV,zerotmp_redner,zerotmp_NV,Janke_zero,Chakraborty2017,Redner2001,Das2017}.

For a more general study, we have carried out the quenching experiments from various temperatures $T_s$ above the critical value. For both the models it appears that systems prepared at higher temperatures in the paramagnetic phase approach the final equilibrium quicker, irrespective of the value of $T_f$. This resembles the Mpemba effect \cite{mpemba_cool}. As opposed to the previous study \cite{vadakkayil,chatterjee}, here we have chosen the initial configurations from the full distributions of order parameter at various $T_s$. 
This suggests that the Mpemba effect during para- to ferromagnetic transitions is of experimental relevance with possible applications.
Interestingly, in the 2D Ising model the effect appears only when a restriction is set on the initial magnetization, viz., $|m|\simeq 0$. 

In an earlier article \cite{chatterjee}, it was argued that differences in critical fluctuations at the initial states are responsible for the Mpemba effect during these magnetic transitions. Careful investigation of the above issue, concerning the space-dimension dependence, can shed important light along this direction. Other important matters are related to finite-size-effects, metastability, etc. While our results suggest that ME can arise without metastability, this matter demands further careful investigation, given that, despite no in-built frustration, large number of runs, e.g., in the Ising model, in $d=2$ as well as $d=3$, encounter freezing \cite{zerotmp_NV}. In future we also plan for a more detailed study with the XY model, particularly in $d=2$ in which this model exhibits Kosterlitz-Thouless \cite{Kosterlitz1973,Kosterlitz1974} transition. Other models exhibiting similar transitions should also be of interest. It should be noted that in recent times quantum Mpemba effect has become a topic of interest \cite{QME_Joshi,QME_Amit,QME_Colin,QME_Zhang,QME_Ares,QME_Moroder,QME_Strachan,QME_XY}. List of works in this domain includes contributions for XY model \cite{QME_XY} as well.

\section*{Author contributions}
SKD designed and supervised the work. WA, NV and SC carried out the work.

\section*{Acknowledgements}
The authors acknowledge computational time in PARAM Yukti supercomputers of National Supercomputing Mission located at JNCASR.

\begin{thebibliography}{77}%
\makeatletter
\providecommand \@ifxundefined [1]{%
 \@ifx{#1\undefined}
}%
\providecommand \@ifnum [1]{%
 \ifnum #1\expandafter \@firstoftwo
 \else \expandafter \@secondoftwo
 \fi
}%
\providecommand \@ifx [1]{%
 \ifx #1\expandafter \@firstoftwo
 \else \expandafter \@secondoftwo
 \fi
}%
\providecommand \natexlab [1]{#1}%
\providecommand \enquote  [1]{``#1''}%
\providecommand \bibnamefont  [1]{#1}%
\providecommand \bibfnamefont [1]{#1}%
\providecommand \citenamefont [1]{#1}%
\providecommand \href@noop [0]{\@secondoftwo}%
\providecommand \href [0]{\begingroup \@sanitize@url \@href}%
\providecommand \@href[1]{\@@startlink{#1}\@@href}%
\providecommand \@@href[1]{\endgroup#1\@@endlink}%
\providecommand \@sanitize@url [0]{\catcode `\\12\catcode `\$12\catcode
  `\&12\catcode `\#12\catcode `\^12\catcode `\_12\catcode `\%12\relax}%
\providecommand \@@startlink[1]{}%
\providecommand \@@endlink[0]{}%
\providecommand \url  [0]{\begingroup\@sanitize@url \@url }%
\providecommand \@url [1]{\endgroup\@href {#1}{\urlprefix }}%
\providecommand \urlprefix  [0]{URL }%
\providecommand \Eprint [0]{\href }%
\providecommand \doibase [0]{https://doi.org/}%
\providecommand \selectlanguage [0]{\@gobble}%
\providecommand \bibinfo  [0]{\@secondoftwo}%
\providecommand \bibfield  [0]{\@secondoftwo}%
\providecommand \translation [1]{[#1]}%
\providecommand \BibitemOpen [0]{}%
\providecommand \bibitemStop [0]{}%
\providecommand \bibitemNoStop [0]{.\EOS\space}%
\providecommand \EOS [0]{\spacefactor3000\relax}%
\providecommand \BibitemShut  [1]{\csname bibitem#1\endcsname}%
\let\auto@bib@innerbib\@empty
\bibitem [{\citenamefont {Mpemba}\ and\ \citenamefont
  {Osborne}(1969)}]{mpemba_cool}%
  \BibitemOpen
  \bibfield  {author} {\bibinfo {author} {\bibfnamefont {E.~B.}\ \bibnamefont
  {Mpemba}}\ and\ \bibinfo {author} {\bibfnamefont {D.~G.}\ \bibnamefont
  {Osborne}},\ }\bibfield  {title} {\bibinfo {title} {Cool?},\ }\href
  {https://doi.org/10.1088/0031-9120/4/3/312} {\bibfield  {journal} {\bibinfo
  {journal} {Phys. Educ.}\ }\textbf {\bibinfo {volume} {4}},\ \bibinfo {pages}
  {172} (\bibinfo {year} {1969})}\BibitemShut {NoStop}%
\bibitem [{\citenamefont {Tang}\ \emph {et~al.}(2023)\citenamefont {Tang},
  \citenamefont {Huang}, \citenamefont {Zhang}, \citenamefont {Liu},\ and\
  \citenamefont {Zhao}}]{Tang2023Mpemba}%
  \BibitemOpen
  \bibfield  {author} {\bibinfo {author} {\bibfnamefont {Z.}~\bibnamefont
  {Tang}}, \bibinfo {author} {\bibfnamefont {W.}~\bibnamefont {Huang}},
  \bibinfo {author} {\bibfnamefont {Y.}~\bibnamefont {Zhang}}, \bibinfo
  {author} {\bibfnamefont {Y.}~\bibnamefont {Liu}},\ and\ \bibinfo {author}
  {\bibfnamefont {L.}~\bibnamefont {Zhao}},\ }\bibfield  {title} {\bibinfo
  {title} {{Direct observation of the Mpemba effect with water: Probe the
  mysterious heat transfer}},\ }\href {https://doi.org/10.1002/inf2.12352}
  {\bibfield  {journal} {\bibinfo  {journal} {InfoMat}\ }\textbf {\bibinfo
  {volume} {5}},\ \bibinfo {pages} {e12352} (\bibinfo {year}
  {2023})}\BibitemShut {NoStop}%
\bibitem [{\citenamefont {Burridge}\ and\ \citenamefont
  {Hallstadius}(2020)}]{Burridge2020Mpemba}%
  \BibitemOpen
  \bibfield  {author} {\bibinfo {author} {\bibfnamefont {H.~C.}\ \bibnamefont
  {Burridge}}\ and\ \bibinfo {author} {\bibfnamefont {O.}~\bibnamefont
  {Hallstadius}},\ }\bibfield  {title} {\bibinfo {title} {{Observing the Mpemba
  effect with minimal bias and the value of the Mpemba effect to scientific
  outreach and engagement}},\ }\href {https://doi.org/10.1098/rspa.2019.0829}
  {\bibfield  {journal} {\bibinfo  {journal} {Proc. R. Soc. A}\ }\textbf
  {\bibinfo {volume} {476}},\ \bibinfo {pages} {20190829} (\bibinfo {year}
  {2020})}\BibitemShut {NoStop}%
\bibitem [{\citenamefont {Ghosh}\ \emph {et~al.}(2025)\citenamefont {Ghosh},
  \citenamefont {Pathak}, \citenamefont {Chatterjee},\ and\ \citenamefont
  {Das}}]{Ghosh2025}%
  \BibitemOpen
  \bibfield  {author} {\bibinfo {author} {\bibfnamefont {S.}~\bibnamefont
  {Ghosh}}, \bibinfo {author} {\bibfnamefont {P.}~\bibnamefont {Pathak}},
  \bibinfo {author} {\bibfnamefont {S.}~\bibnamefont {Chatterjee}},\ and\
  \bibinfo {author} {\bibfnamefont {S.~K.}\ \bibnamefont {Das}},\ }\bibfield
  {title} {\bibinfo {title} {{Simulations of Mpemba effect in water and
  Lennard-Jones models}},\ }\href {https://doi.org/10.1038/s42005-025-02251-6}
  {\bibfield  {journal} {\bibinfo  {journal} {Commun. Phys.}\ }\textbf
  {\bibinfo {volume} {8}},\ \bibinfo {pages} {359} (\bibinfo {year}
  {2025})}\BibitemShut {NoStop}%
\bibitem [{\citenamefont {Teza}\ \emph {et~al.}(2026)\citenamefont {Teza},
  \citenamefont {Bechhoefer}, \citenamefont {Lasanta}, \citenamefont {Raz},\
  and\ \citenamefont {Vucelja}}]{Teza2026MpembaReview}%
  \BibitemOpen
  \bibfield  {author} {\bibinfo {author} {\bibfnamefont {G.}~\bibnamefont
  {Teza}}, \bibinfo {author} {\bibfnamefont {J.}~\bibnamefont {Bechhoefer}},
  \bibinfo {author} {\bibfnamefont {A.}~\bibnamefont {Lasanta}}, \bibinfo
  {author} {\bibfnamefont {O.}~\bibnamefont {Raz}},\ and\ \bibinfo {author}
  {\bibfnamefont {M.}~\bibnamefont {Vucelja}},\ }\bibfield  {title} {\bibinfo
  {title} {Speedups in nonequilibrium thermal relaxation: {Mpemba} and related
  effects},\ }\href {https://doi.org/10.1016/j.physrep.2025.10.009} {\bibfield
  {journal} {\bibinfo  {journal} {Phys. Rep.}\ }\textbf {\bibinfo {volume}
  {1164}},\ \bibinfo {pages} {1} (\bibinfo {year} {2026})}\BibitemShut
  {NoStop}%
\bibitem [{\citenamefont {Bechhoefer}\ \emph {et~al.}(2021)\citenamefont
  {Bechhoefer}, \citenamefont {Kumar},\ and\ \citenamefont
  {Ch\'etrite}}]{Bechhoefer2021}%
  \BibitemOpen
  \bibfield  {author} {\bibinfo {author} {\bibfnamefont {J.}~\bibnamefont
  {Bechhoefer}}, \bibinfo {author} {\bibfnamefont {A.}~\bibnamefont {Kumar}},\
  and\ \bibinfo {author} {\bibfnamefont {R.}~\bibnamefont {Ch\'etrite}},\
  }\bibfield  {title} {\bibinfo {title} {{A fresh understanding of the Mpemba
  effect}},\ }\href {https://doi.org/10.1038/s42254-021-00349-8} {\bibfield
  {journal} {\bibinfo  {journal} {Nat. Rev. Phys.}\ }\textbf {\bibinfo {volume}
  {3}},\ \bibinfo {pages} {534} (\bibinfo {year} {2021})}\BibitemShut {NoStop}%
\bibitem [{\citenamefont {Auerbach}(1995)}]{Auerbach1995}%
  \BibitemOpen
  \bibfield  {author} {\bibinfo {author} {\bibfnamefont {D.}~\bibnamefont
  {Auerbach}},\ }\bibfield  {title} {\bibinfo {title} {{Supercooling and the
  Mpemba effect: When hot water freezes quicker than cold}},\ }\href
  {https://doi.org/10.1119/1.18059} {\bibfield  {journal} {\bibinfo  {journal}
  {Am. J. Phys.}\ }\textbf {\bibinfo {volume} {63}},\ \bibinfo {pages} {882}
  (\bibinfo {year} {1995})}\BibitemShut {NoStop}%
\bibitem [{\citenamefont {Das}(2023)}]{langmuir_skd}%
  \BibitemOpen
  \bibfield  {author} {\bibinfo {author} {\bibfnamefont {S.~K.}\ \bibnamefont
  {Das}},\ }\bibfield  {title} {\bibinfo {title} {{Perspectives on a Few
  Puzzles in Phase Transformations: When Should the Farthest Reach the
  Earliest?}},\ }\href {https://doi.org/10.1021/acs.langmuir.3c00668}
  {\bibfield  {journal} {\bibinfo  {journal} {Langmuir}\ }\textbf {\bibinfo
  {volume} {39}},\ \bibinfo {pages} {10715} (\bibinfo {year}
  {2023})}\BibitemShut {NoStop}%
\bibitem [{\citenamefont {Jeng}(2006)}]{Jeng2006}%
  \BibitemOpen
  \bibfield  {author} {\bibinfo {author} {\bibfnamefont {M.}~\bibnamefont
  {Jeng}},\ }\bibfield  {title} {\bibinfo {title} {{The Mpemba effect: When can
  hot water freeze faster than cold?}},\ }\href
  {https://doi.org/10.1119/1.2186331} {\bibfield  {journal} {\bibinfo
  {journal} {Am. J. Phys.}\ }\textbf {\bibinfo {volume} {74}},\ \bibinfo
  {pages} {514} (\bibinfo {year} {2006})}\BibitemShut {NoStop}%
\bibitem [{\citenamefont {Lu}\ and\ \citenamefont {Raz}(2017)}]{Oraz}%
  \BibitemOpen
  \bibfield  {author} {\bibinfo {author} {\bibfnamefont {Z.}~\bibnamefont
  {Lu}}\ and\ \bibinfo {author} {\bibfnamefont {O.}~\bibnamefont {Raz}},\
  }\bibfield  {title} {\bibinfo {title} {{Nonequilibrium thermodynamics of the
  Markovian Mpemba effect and its inverse}},\ }\href
  {https://doi.org/10.1073/pnas.1701264114} {\bibfield  {journal} {\bibinfo
  {journal} {Proc. Natl. Acad. Sci. U.S.A.}\ }\textbf {\bibinfo {volume}
  {114}},\ \bibinfo {pages} {5083} (\bibinfo {year} {2017})}\BibitemShut
  {NoStop}%
\bibitem [{\citenamefont {Klich}\ \emph {et~al.}(2019)\citenamefont {Klich},
  \citenamefont {Raz}, \citenamefont {Hirschberg},\ and\ \citenamefont
  {Vucelja}}]{Oraz_prx}%
  \BibitemOpen
  \bibfield  {author} {\bibinfo {author} {\bibfnamefont {I.}~\bibnamefont
  {Klich}}, \bibinfo {author} {\bibfnamefont {O.}~\bibnamefont {Raz}}, \bibinfo
  {author} {\bibfnamefont {O.}~\bibnamefont {Hirschberg}},\ and\ \bibinfo
  {author} {\bibfnamefont {M.}~\bibnamefont {Vucelja}},\ }\bibfield  {title}
  {\bibinfo {title} {{Mpemba Index and Anomalous Relaxation}},\ }\href
  {https://doi.org/10.1103/PhysRevX.9.021060} {\bibfield  {journal} {\bibinfo
  {journal} {Phys. Rev. X}\ }\textbf {\bibinfo {volume} {9}},\ \bibinfo {pages}
  {021060} (\bibinfo {year} {2019})}\BibitemShut {NoStop}%
\bibitem [{\citenamefont {Kumar}\ and\ \citenamefont
  {Bechhoefer}(2020)}]{avinash}%
  \BibitemOpen
  \bibfield  {author} {\bibinfo {author} {\bibfnamefont {A.}~\bibnamefont
  {Kumar}}\ and\ \bibinfo {author} {\bibfnamefont {J.}~\bibnamefont
  {Bechhoefer}},\ }\bibfield  {title} {\bibinfo {title} {{Exponentially faster
  cooling in a colloidal system}},\ }\href
  {https://doi.org/https://doi.org/10.1038/s41586-020-2560-x} {\bibfield
  {journal} {\bibinfo  {journal} {Nature}\ }\textbf {\bibinfo {volume} {584}},\
  \bibinfo {pages} {64} (\bibinfo {year} {2020})}\BibitemShut {NoStop}%
\bibitem [{\citenamefont {Ch{\'e}trite}\ \emph {et~al.}(2021)\citenamefont
  {Ch{\'e}trite}, \citenamefont {Kumar},\ and\ \citenamefont
  {Bechhoefer}}]{chetrite2021metastable}%
  \BibitemOpen
  \bibfield  {author} {\bibinfo {author} {\bibfnamefont {R.}~\bibnamefont
  {Ch{\'e}trite}}, \bibinfo {author} {\bibfnamefont {A.}~\bibnamefont
  {Kumar}},\ and\ \bibinfo {author} {\bibfnamefont {J.}~\bibnamefont
  {Bechhoefer}},\ }\bibfield  {title} {\bibinfo {title} {{The metastable
  {M}pemba effect corresponds to a non-monotonic temperature dependence of
  extractable work}},\ }\href {https://doi.org/10.3389/fphy.2021.654271}
  {\bibfield  {journal} {\bibinfo  {journal} {Front. Phys.}\ }\textbf {\bibinfo
  {volume} {9}},\ \bibinfo {pages} {654271} (\bibinfo {year}
  {2021})}\BibitemShut {NoStop}%
\bibitem [{\citenamefont {Schwarzendahl}\ and\ \citenamefont
  {L\"owen}(2022)}]{Hartmut}%
  \BibitemOpen
  \bibfield  {author} {\bibinfo {author} {\bibfnamefont {F.~J.}\ \bibnamefont
  {Schwarzendahl}}\ and\ \bibinfo {author} {\bibfnamefont {H.}~\bibnamefont
  {L\"owen}},\ }\bibfield  {title} {\bibinfo {title} {Anomalous cooling and
  overcooling of active colloids},\ }\href
  {https://doi.org/10.1103/PhysRevLett.129.138002} {\bibfield  {journal}
  {\bibinfo  {journal} {Phys. Rev. Lett.}\ }\textbf {\bibinfo {volume} {129}},\
  \bibinfo {pages} {138002} (\bibinfo {year} {2022})}\BibitemShut {NoStop}%
\bibitem [{\citenamefont {Vadakkayil}\ and\ \citenamefont
  {Das}(2021)}]{vadakkayil}%
  \BibitemOpen
  \bibfield  {author} {\bibinfo {author} {\bibfnamefont {N.}~\bibnamefont
  {Vadakkayil}}\ and\ \bibinfo {author} {\bibfnamefont {S.~K.}\ \bibnamefont
  {Das}},\ }\bibfield  {title} {\bibinfo {title} {{Should a hotter paramagnet
  transform quicker to a ferromagnet? Monte Carlo simulation results for Ising
  model}},\ }\href {https://doi.org/10.1039/D1CP00879J} {\bibfield  {journal}
  {\bibinfo  {journal} {Phys. Chem. Chem. Phys.}\ }\textbf {\bibinfo {volume}
  {23}},\ \bibinfo {pages} {11186} (\bibinfo {year} {2021})}\BibitemShut
  {NoStop}%
\bibitem [{\citenamefont {Chatterjee}\ \emph {et~al.}(2024)\citenamefont
  {Chatterjee}, \citenamefont {Ghosh}, \citenamefont {Vadakkayil},
  \citenamefont {Paul}, \citenamefont {Singha},\ and\ \citenamefont
  {Das}}]{chatterjee}%
  \BibitemOpen
  \bibfield  {author} {\bibinfo {author} {\bibfnamefont {S.}~\bibnamefont
  {Chatterjee}}, \bibinfo {author} {\bibfnamefont {S.}~\bibnamefont {Ghosh}},
  \bibinfo {author} {\bibfnamefont {N.}~\bibnamefont {Vadakkayil}}, \bibinfo
  {author} {\bibfnamefont {T.}~\bibnamefont {Paul}}, \bibinfo {author}
  {\bibfnamefont {S.~K.}\ \bibnamefont {Singha}},\ and\ \bibinfo {author}
  {\bibfnamefont {S.~K.}\ \bibnamefont {Das}},\ }\bibfield  {title} {\bibinfo
  {title} {{Mpemba effect in pure spin systems : A universal picture of the
  role of spatial correlations at initial states}},\ }\href
  {https://doi.org/10.1103/PhysRevE.110.L012103} {\bibfield  {journal}
  {\bibinfo  {journal} {Phys. Rev. E}\ }\textbf {\bibinfo {volume} {110}},\
  \bibinfo {pages} {L012103} (\bibinfo {year} {2024})}\BibitemShut {NoStop}%
\bibitem [{\citenamefont {Chatterjee}\ \emph {et~al.}(2025)\citenamefont
  {Chatterjee}, \citenamefont {Das}, \citenamefont {Pathak}, \citenamefont
  {Paul},\ and\ \citenamefont {Das}}]{sohom_vicsek}%
  \BibitemOpen
  \bibfield  {author} {\bibinfo {author} {\bibfnamefont {S.}~\bibnamefont
  {Chatterjee}}, \bibinfo {author} {\bibfnamefont {S.}~\bibnamefont {Das}},
  \bibinfo {author} {\bibfnamefont {P.}~\bibnamefont {Pathak}}, \bibinfo
  {author} {\bibfnamefont {T.}~\bibnamefont {Paul}},\ and\ \bibinfo {author}
  {\bibfnamefont {S.~K.}\ \bibnamefont {Das}},\ }\bibfield  {title} {\bibinfo
  {title} {{Quicker flocking in aligning active matters for noisier
  beginning}},\ }\href {https://arxiv.org/abs/2502.06482} {\bibfield  {journal}
  {\bibinfo  {journal} {arXiv: 2502.06482}\ } (\bibinfo {year}
  {2025})}\BibitemShut {NoStop}%
\bibitem [{\citenamefont {Lasanta}\ \emph {et~al.}(2017)\citenamefont
  {Lasanta}, \citenamefont {Vega~Reyes}, \citenamefont {Prados},\ and\
  \citenamefont {Santos}}]{granular_lasanta}%
  \BibitemOpen
  \bibfield  {author} {\bibinfo {author} {\bibfnamefont {A.}~\bibnamefont
  {Lasanta}}, \bibinfo {author} {\bibfnamefont {F.}~\bibnamefont {Vega~Reyes}},
  \bibinfo {author} {\bibfnamefont {A.}~\bibnamefont {Prados}},\ and\ \bibinfo
  {author} {\bibfnamefont {A.}~\bibnamefont {Santos}},\ }\bibfield  {title}
  {\bibinfo {title} {{When the Hotter Cools More Quickly: Mpemba Effect in
  Granular Fluids}},\ }\href {https://doi.org/10.1103/PhysRevLett.119.148001}
  {\bibfield  {journal} {\bibinfo  {journal} {Phys. Rev. Lett.}\ }\textbf
  {\bibinfo {volume} {119}},\ \bibinfo {pages} {148001} (\bibinfo {year}
  {2017})}\BibitemShut {NoStop}%
\bibitem [{\citenamefont {Biswas}\ \emph
  {et~al.}(2023{\natexlab{a}})\citenamefont {Biswas}, \citenamefont {Prasad},\
  and\ \citenamefont {Rajesh}}]{granular_biswas}%
  \BibitemOpen
  \bibfield  {author} {\bibinfo {author} {\bibfnamefont {A.}~\bibnamefont
  {Biswas}}, \bibinfo {author} {\bibfnamefont {V.~V.}\ \bibnamefont {Prasad}},\
  and\ \bibinfo {author} {\bibfnamefont {R.}~\bibnamefont {Rajesh}},\
  }\bibfield  {title} {\bibinfo {title} {Mpemba effect in driven granular
  gases: Role of distance measures},\ }\href
  {https://doi.org/10.1103/PhysRevE.108.024902} {\bibfield  {journal} {\bibinfo
   {journal} {Phys. Rev. E}\ }\textbf {\bibinfo {volume} {108}},\ \bibinfo
  {pages} {024902} (\bibinfo {year} {2023}{\natexlab{a}})}\BibitemShut
  {NoStop}%
\bibitem [{\citenamefont {Chaddah}\ \emph {et~al.}(2010)\citenamefont
  {Chaddah}, \citenamefont {Dash}, \citenamefont {Kumar},\ and\ \citenamefont
  {Banerjee}}]{Chaddah2010}%
  \BibitemOpen
  \bibfield  {author} {\bibinfo {author} {\bibfnamefont {P.}~\bibnamefont
  {Chaddah}}, \bibinfo {author} {\bibfnamefont {S.}~\bibnamefont {Dash}},
  \bibinfo {author} {\bibfnamefont {K.}~\bibnamefont {Kumar}},\ and\ \bibinfo
  {author} {\bibfnamefont {A.}~\bibnamefont {Banerjee}},\ }\bibfield  {title}
  {\bibinfo {title} {Overtaking while approaching equilibrium},\ }\href
  {https://arxiv.org/abs/1011.3598} {\bibfield  {journal} {\bibinfo  {journal}
  {arXiv: 1011.3598}\ } (\bibinfo {year} {2010})}\BibitemShut {NoStop}%
\bibitem [{\citenamefont {Greaney}\ \emph {et~al.}(2011)\citenamefont
  {Greaney}, \citenamefont {Lani}, \citenamefont {Cicero},\ and\ \citenamefont
  {Grossman}}]{Carbon_nano2011}%
  \BibitemOpen
  \bibfield  {author} {\bibinfo {author} {\bibfnamefont {P.~A.}\ \bibnamefont
  {Greaney}}, \bibinfo {author} {\bibfnamefont {G.}~\bibnamefont {Lani}},
  \bibinfo {author} {\bibfnamefont {G.}~\bibnamefont {Cicero}},\ and\ \bibinfo
  {author} {\bibfnamefont {J.~C.}\ \bibnamefont {Grossman}},\ }\bibfield
  {title} {\bibinfo {title} {{Mpemba-Like Behavior in Carbon Nanotube
  Resonators}},\ }\href {https://doi.org/10.1007/s11661-011-0843-4} {\bibfield
  {journal} {\bibinfo  {journal} {Metall. Mater. Trans. A}\ }\textbf {\bibinfo
  {volume} {42}},\ \bibinfo {pages} {3907} (\bibinfo {year}
  {2011})}\BibitemShut {NoStop}%
\bibitem [{\citenamefont {Ahn}\ \emph {et~al.}(2016)\citenamefont {Ahn},
  \citenamefont {Kang}, \citenamefont {Koh},\ and\ \citenamefont
  {Lee}}]{clathrate2016}%
  \BibitemOpen
  \bibfield  {author} {\bibinfo {author} {\bibfnamefont {Y.-H.}\ \bibnamefont
  {Ahn}}, \bibinfo {author} {\bibfnamefont {H.}~\bibnamefont {Kang}}, \bibinfo
  {author} {\bibfnamefont {D.-Y.}\ \bibnamefont {Koh}},\ and\ \bibinfo {author}
  {\bibfnamefont {H.}~\bibnamefont {Lee}},\ }\bibfield  {title} {\bibinfo
  {title} {{Experimental verifications of Mpemba-like behaviors of clathrate
  hydrates}},\ }\href {https://doi.org/10.1007/s11814-016-0029-2} {\bibfield
  {journal} {\bibinfo  {journal} {Korean J. Chem. Eng.}\ }\textbf {\bibinfo
  {volume} {33}},\ \bibinfo {pages} {1903} (\bibinfo {year}
  {2016})}\BibitemShut {NoStop}%
\bibitem [{\citenamefont {Van~Vu}\ and\ \citenamefont
  {Hayakawa}(2025)}]{Thermomajorization}%
  \BibitemOpen
  \bibfield  {author} {\bibinfo {author} {\bibfnamefont {T.}~\bibnamefont
  {Van~Vu}}\ and\ \bibinfo {author} {\bibfnamefont {H.}~\bibnamefont
  {Hayakawa}},\ }\bibfield  {title} {\bibinfo {title} {{Thermomajorization
  Mpemba Effect}},\ }\href {https://doi.org/10.1103/PhysRevLett.134.107101}
  {\bibfield  {journal} {\bibinfo  {journal} {Phys. Rev. Lett.}\ }\textbf
  {\bibinfo {volume} {134}},\ \bibinfo {pages} {107101} (\bibinfo {year}
  {2025})}\BibitemShut {NoStop}%
\bibitem [{\citenamefont {Tian}\ \emph {et~al.}(2025)\citenamefont {Tian},
  \citenamefont {Zheng}, \citenamefont {Liu}, \citenamefont {Wang},
  \citenamefont {Guo},\ and\ \citenamefont {Sun}}]{Langevin_exp}%
  \BibitemOpen
  \bibfield  {author} {\bibinfo {author} {\bibfnamefont {Y.}~\bibnamefont
  {Tian}}, \bibinfo {author} {\bibfnamefont {Y.}~\bibnamefont {Zheng}},
  \bibinfo {author} {\bibfnamefont {L.-H.}\ \bibnamefont {Liu}}, \bibinfo
  {author} {\bibfnamefont {L.}~\bibnamefont {Wang}}, \bibinfo {author}
  {\bibfnamefont {G.-C.}\ \bibnamefont {Guo}},\ and\ \bibinfo {author}
  {\bibfnamefont {F.-W.}\ \bibnamefont {Sun}},\ }\bibfield  {title} {\bibinfo
  {title} {{Experimental study of Mpemba effect in an energy Langevin
  system}},\ }\href {https://doi.org/10.1103/5p4l-1515} {\bibfield  {journal}
  {\bibinfo  {journal} {Phys. Rev. Res.}\ }\textbf {\bibinfo {volume} {7}},\
  \bibinfo {pages} {L042020} (\bibinfo {year} {2025})}\BibitemShut {NoStop}%
\bibitem [{\citenamefont {Joshi}\ \emph {et~al.}(2024)\citenamefont {Joshi},
  \citenamefont {Franke}, \citenamefont {Rath}, \citenamefont {Ares},
  \citenamefont {Murciano}, \citenamefont {Kranzl}, \citenamefont {Blatt},
  \citenamefont {Zoller}, \citenamefont {Vermersch}, \citenamefont {Calabrese},
  \citenamefont {Roos},\ and\ \citenamefont {Joshi}}]{QME_Joshi}%
  \BibitemOpen
  \bibfield  {author} {\bibinfo {author} {\bibfnamefont {L.~K.}\ \bibnamefont
  {Joshi}}, \bibinfo {author} {\bibfnamefont {J.}~\bibnamefont {Franke}},
  \bibinfo {author} {\bibfnamefont {A.}~\bibnamefont {Rath}}, \bibinfo {author}
  {\bibfnamefont {F.}~\bibnamefont {Ares}}, \bibinfo {author} {\bibfnamefont
  {S.}~\bibnamefont {Murciano}}, \bibinfo {author} {\bibfnamefont
  {F.}~\bibnamefont {Kranzl}}, \bibinfo {author} {\bibfnamefont
  {R.}~\bibnamefont {Blatt}}, \bibinfo {author} {\bibfnamefont
  {P.}~\bibnamefont {Zoller}}, \bibinfo {author} {\bibfnamefont
  {B.}~\bibnamefont {Vermersch}}, \bibinfo {author} {\bibfnamefont
  {P.}~\bibnamefont {Calabrese}}, \bibinfo {author} {\bibfnamefont {C.~F.}\
  \bibnamefont {Roos}},\ and\ \bibinfo {author} {\bibfnamefont {M.~K.}\
  \bibnamefont {Joshi}},\ }\bibfield  {title} {\bibinfo {title} {{Observing the
  Quantum Mpemba Effect in Quantum Simulations}},\ }\href
  {https://doi.org/10.1103/PhysRevLett.133.010402} {\bibfield  {journal}
  {\bibinfo  {journal} {Phys. Rev. Lett.}\ }\textbf {\bibinfo {volume} {133}},\
  \bibinfo {pages} {010402} (\bibinfo {year} {2024})}\BibitemShut {NoStop}%
\bibitem [{\citenamefont {Chatterjee}\ \emph {et~al.}(2023)\citenamefont
  {Chatterjee}, \citenamefont {Takada},\ and\ \citenamefont
  {Hayakawa}}]{QME_Amit}%
  \BibitemOpen
  \bibfield  {author} {\bibinfo {author} {\bibfnamefont {A.~K.}\ \bibnamefont
  {Chatterjee}}, \bibinfo {author} {\bibfnamefont {S.}~\bibnamefont {Takada}},\
  and\ \bibinfo {author} {\bibfnamefont {H.}~\bibnamefont {Hayakawa}},\
  }\bibfield  {title} {\bibinfo {title} {{Quantum Mpemba Effect in a Quantum
  Dot with Reservoirs}},\ }\href
  {https://doi.org/10.1103/PhysRevLett.131.080402} {\bibfield  {journal}
  {\bibinfo  {journal} {Phys. Rev. Lett.}\ }\textbf {\bibinfo {volume} {131}},\
  \bibinfo {pages} {080402} (\bibinfo {year} {2023})}\BibitemShut {NoStop}%
\bibitem [{\citenamefont {Rylands}\ \emph {et~al.}(2024)\citenamefont
  {Rylands}, \citenamefont {Klobas}, \citenamefont {Ares}, \citenamefont
  {Calabrese}, \citenamefont {Murciano},\ and\ \citenamefont
  {Bertini}}]{QME_Colin}%
  \BibitemOpen
  \bibfield  {author} {\bibinfo {author} {\bibfnamefont {C.}~\bibnamefont
  {Rylands}}, \bibinfo {author} {\bibfnamefont {K.}~\bibnamefont {Klobas}},
  \bibinfo {author} {\bibfnamefont {F.}~\bibnamefont {Ares}}, \bibinfo {author}
  {\bibfnamefont {P.}~\bibnamefont {Calabrese}}, \bibinfo {author}
  {\bibfnamefont {S.}~\bibnamefont {Murciano}},\ and\ \bibinfo {author}
  {\bibfnamefont {B.}~\bibnamefont {Bertini}},\ }\bibfield  {title} {\bibinfo
  {title} {{Microscopic Origin of the Quantum Mpemba Effect in Integrable
  Systems}},\ }\href {https://doi.org/10.1103/PhysRevLett.133.010401}
  {\bibfield  {journal} {\bibinfo  {journal} {Phys. Rev. Lett.}\ }\textbf
  {\bibinfo {volume} {133}},\ \bibinfo {pages} {010401} (\bibinfo {year}
  {2024})}\BibitemShut {NoStop}%
\bibitem [{\citenamefont {Zhang}\ \emph {et~al.}(2025)\citenamefont {Zhang},
  \citenamefont {Xia}, \citenamefont {Wu}, \citenamefont {Chen}, \citenamefont
  {Zhang}, \citenamefont {Xie}, \citenamefont {Su}, \citenamefont {Wu},
  \citenamefont {Qiu}, \citenamefont {Chen}, \citenamefont {Li}, \citenamefont
  {Jing},\ and\ \citenamefont {Zhou}}]{QME_Zhang}%
  \BibitemOpen
  \bibfield  {author} {\bibinfo {author} {\bibfnamefont {J.}~\bibnamefont
  {Zhang}}, \bibinfo {author} {\bibfnamefont {G.}~\bibnamefont {Xia}}, \bibinfo
  {author} {\bibfnamefont {C.-W.}\ \bibnamefont {Wu}}, \bibinfo {author}
  {\bibfnamefont {T.}~\bibnamefont {Chen}}, \bibinfo {author} {\bibfnamefont
  {Q.}~\bibnamefont {Zhang}}, \bibinfo {author} {\bibfnamefont
  {Y.}~\bibnamefont {Xie}}, \bibinfo {author} {\bibfnamefont {W.-B.}\
  \bibnamefont {Su}}, \bibinfo {author} {\bibfnamefont {W.}~\bibnamefont {Wu}},
  \bibinfo {author} {\bibfnamefont {C.-W.}\ \bibnamefont {Qiu}}, \bibinfo
  {author} {\bibfnamefont {P.-X.}\ \bibnamefont {Chen}}, \bibinfo {author}
  {\bibfnamefont {W.}~\bibnamefont {Li}}, \bibinfo {author} {\bibfnamefont
  {H.}~\bibnamefont {Jing}},\ and\ \bibinfo {author} {\bibfnamefont {Y.-L.}\
  \bibnamefont {Zhou}},\ }\bibfield  {title} {\bibinfo {title} {{Observation of
  quantum strong Mpemba effect}},\ }\href
  {https://doi.org/10.1038/s41467-024-54303-0} {\bibfield  {journal} {\bibinfo
  {journal} {Nat. Commun.}\ }\textbf {\bibinfo {volume} {16}},\ \bibinfo
  {pages} {301} (\bibinfo {year} {2025})}\BibitemShut {NoStop}%
\bibitem [{\citenamefont {Ares}\ \emph {et~al.}(2025)\citenamefont {Ares},
  \citenamefont {Calabrese},\ and\ \citenamefont {Murciano}}]{QME_Ares}%
  \BibitemOpen
  \bibfield  {author} {\bibinfo {author} {\bibfnamefont {F.}~\bibnamefont
  {Ares}}, \bibinfo {author} {\bibfnamefont {P.}~\bibnamefont {Calabrese}},\
  and\ \bibinfo {author} {\bibfnamefont {S.}~\bibnamefont {Murciano}},\
  }\bibfield  {title} {\bibinfo {title} {{The quantum Mpemba effects}},\ }\href
  {https://doi.org/10.1038/s42254-025-00838-0} {\bibfield  {journal} {\bibinfo
  {journal} {Nat. Rev. Phys.}\ }\textbf {\bibinfo {volume} {7}},\ \bibinfo
  {pages} {451} (\bibinfo {year} {2025})}\BibitemShut {NoStop}%
\bibitem [{\citenamefont {Moroder}\ \emph {et~al.}(2024)\citenamefont
  {Moroder}, \citenamefont {Culhane}, \citenamefont {Zawadzki},\ and\
  \citenamefont {Goold}}]{QME_Moroder}%
  \BibitemOpen
  \bibfield  {author} {\bibinfo {author} {\bibfnamefont {M.}~\bibnamefont
  {Moroder}}, \bibinfo {author} {\bibfnamefont {O.}~\bibnamefont {Culhane}},
  \bibinfo {author} {\bibfnamefont {K.}~\bibnamefont {Zawadzki}},\ and\
  \bibinfo {author} {\bibfnamefont {J.}~\bibnamefont {Goold}},\ }\bibfield
  {title} {\bibinfo {title} {{Thermodynamics of the Quantum Mpemba Effect}},\
  }\href {https://doi.org/10.1103/PhysRevLett.133.140404} {\bibfield  {journal}
  {\bibinfo  {journal} {Phys. Rev. Lett.}\ }\textbf {\bibinfo {volume} {133}},\
  \bibinfo {pages} {140404} (\bibinfo {year} {2024})}\BibitemShut {NoStop}%
\bibitem [{\citenamefont {Strachan}\ \emph {et~al.}(2025)\citenamefont
  {Strachan}, \citenamefont {Purkayastha},\ and\ \citenamefont
  {Clark}}]{QME_Strachan}%
  \BibitemOpen
  \bibfield  {author} {\bibinfo {author} {\bibfnamefont {D.~J.}\ \bibnamefont
  {Strachan}}, \bibinfo {author} {\bibfnamefont {A.}~\bibnamefont
  {Purkayastha}},\ and\ \bibinfo {author} {\bibfnamefont {S.~R.}\ \bibnamefont
  {Clark}},\ }\bibfield  {title} {\bibinfo {title} {{Non-Markovian Quantum
  Mpemba Effect}},\ }\href {https://doi.org/10.1103/PhysRevLett.134.220403}
  {\bibfield  {journal} {\bibinfo  {journal} {Phys. Rev. Lett.}\ }\textbf
  {\bibinfo {volume} {134}},\ \bibinfo {pages} {220403} (\bibinfo {year}
  {2025})}\BibitemShut {NoStop}%
\bibitem [{\citenamefont {Murciano}\ \emph {et~al.}(2024)\citenamefont
  {Murciano}, \citenamefont {Ares}, \citenamefont {Klich},\ and\ \citenamefont
  {Calabrese}}]{QME_XY}%
  \BibitemOpen
  \bibfield  {author} {\bibinfo {author} {\bibfnamefont {S.}~\bibnamefont
  {Murciano}}, \bibinfo {author} {\bibfnamefont {F.}~\bibnamefont {Ares}},
  \bibinfo {author} {\bibfnamefont {I.}~\bibnamefont {Klich}},\ and\ \bibinfo
  {author} {\bibfnamefont {P.}~\bibnamefont {Calabrese}},\ }\bibfield  {title}
  {\bibinfo {title} {{Entanglement asymmetry and quantum Mpemba effect in the
  {XY} spin chain}},\ }\href {https://doi.org/10.1088/1742-5468/ad17b4}
  {\bibfield  {journal} {\bibinfo  {journal} {J. Stat. Mech.: Theory Exp.}\
  }\textbf {\bibinfo {volume} {2024}}\bibinfo  {number} { (1)},\ \bibinfo
  {pages} {013103}}\BibitemShut {NoStop}%
\bibitem [{\citenamefont {{M. Baity-Jesi et. al.}}(2019)}]{baity}%
  \BibitemOpen
\bibfield  {number} {  }\bibfield  {author} {\bibinfo {author} {\bibnamefont
  {{M. Baity-Jesi et. al.}}},\ }\bibfield  {title} {\bibinfo {title} {{The
  Mpemba effect in spin glasses is a persistent memory effect}},\ }\href
  {https://doi.org/10.1073/pnas.1819803116} {\bibfield  {journal} {\bibinfo
  {journal} {Proc. Natl. Acad. Sci. U.S.A.}\ }\textbf {\bibinfo {volume}
  {116}},\ \bibinfo {pages} {15350} (\bibinfo {year} {2019})}\BibitemShut
  {NoStop}%
\bibitem [{\citenamefont {Jin}\ and\ \citenamefont {Goddard}(2015)}]{Jin2015}%
  \BibitemOpen
  \bibfield  {author} {\bibinfo {author} {\bibfnamefont {J.}~\bibnamefont
  {Jin}}\ and\ \bibinfo {author} {\bibfnamefont {W.~A.~I.}\ \bibnamefont
  {Goddard}},\ }\bibfield  {title} {\bibinfo {title} {{Mechanisms Underlying
  the Mpemba Effect in Water from Molecular Dynamics Simulations}},\ }\href
  {https://doi.org/10.1021/jp511752n} {\bibfield  {journal} {\bibinfo
  {journal} {J. Phys. Chem. C}\ }\textbf {\bibinfo {volume} {119}},\ \bibinfo
  {pages} {2622} (\bibinfo {year} {2015})}\BibitemShut {NoStop}%
\bibitem [{\citenamefont {Gij\'on}\ \emph {et~al.}(2019)\citenamefont
  {Gij\'on}, \citenamefont {Lasanta},\ and\ \citenamefont
  {Hern\'andez}}]{Lasanta_water}%
  \BibitemOpen
  \bibfield  {author} {\bibinfo {author} {\bibfnamefont {A.}~\bibnamefont
  {Gij\'on}}, \bibinfo {author} {\bibfnamefont {A.}~\bibnamefont {Lasanta}},\
  and\ \bibinfo {author} {\bibfnamefont {E.~R.}\ \bibnamefont {Hern\'andez}},\
  }\bibfield  {title} {\bibinfo {title} {Paths towards equilibrium in molecular
  systems: The case of water},\ }\href
  {https://doi.org/10.1103/PhysRevE.100.032103} {\bibfield  {journal} {\bibinfo
   {journal} {Phys. Rev. E}\ }\textbf {\bibinfo {volume} {100}},\ \bibinfo
  {pages} {032103} (\bibinfo {year} {2019})}\BibitemShut {NoStop}%
\bibitem [{\citenamefont {Biswas}\ \emph
  {et~al.}(2023{\natexlab{b}})\citenamefont {Biswas}, \citenamefont {Rajesh},\
  and\ \citenamefont {Pal}}]{biswas2023mpemba}%
  \BibitemOpen
  \bibfield  {author} {\bibinfo {author} {\bibfnamefont {A.}~\bibnamefont
  {Biswas}}, \bibinfo {author} {\bibfnamefont {R.}~\bibnamefont {Rajesh}},\
  and\ \bibinfo {author} {\bibfnamefont {A.}~\bibnamefont {Pal}},\ }\bibfield
  {title} {\bibinfo {title} {{Mpemba effect in a Langevin system: Population
  statistics, metastability, and other exact results}},\ }\href
  {https://doi.org/10.1063/5.0155855} {\bibfield  {journal} {\bibinfo
  {journal} {J. Chem. Phys.}\ }\textbf {\bibinfo {volume} {159}},\ \bibinfo
  {pages} {044120} (\bibinfo {year} {2023}{\natexlab{b}})}\BibitemShut
  {NoStop}%
\bibitem [{\citenamefont {Landau}\ and\ \citenamefont
  {Binder}(2021)}]{landau_binder}%
  \BibitemOpen
  \bibfield  {author} {\bibinfo {author} {\bibfnamefont {D.~P.}\ \bibnamefont
  {Landau}}\ and\ \bibinfo {author} {\bibfnamefont {K.}~\bibnamefont
  {Binder}},\ }\href@noop {} {\emph {\bibinfo {title} {{A guide to Monte Carlo
  simulations in statistical physics}}}}\ (\bibinfo  {publisher} {Cambridge
  university press},\ \bibinfo {year} {2021})\BibitemShut {NoStop}%
\bibitem [{\citenamefont {Newman}\ and\ \citenamefont
  {Barkema}(1999)}]{newman1999}%
  \BibitemOpen
  \bibfield  {author} {\bibinfo {author} {\bibfnamefont {M.~E.}\ \bibnamefont
  {Newman}}\ and\ \bibinfo {author} {\bibfnamefont {G.~T.}\ \bibnamefont
  {Barkema}},\ }\href@noop {} {\emph {\bibinfo {title} {{Monte Carlo methods in
  statistical physics}}}}\ (\bibinfo  {publisher} {Clarendon Press},\ \bibinfo
  {year} {1999})\BibitemShut {NoStop}%
\bibitem [{\citenamefont {Binder}\ and\ \citenamefont
  {Heermann}(1992)}]{binder1992}%
  \BibitemOpen
  \bibfield  {author} {\bibinfo {author} {\bibfnamefont {K.}~\bibnamefont
  {Binder}}\ and\ \bibinfo {author} {\bibfnamefont {D.~W.}\ \bibnamefont
  {Heermann}},\ }\href@noop {} {\emph {\bibinfo {title} {{Monte Carlo
  simulation in statistical physics}}}},\ Vol.~\bibinfo {volume} {8}\ (\bibinfo
   {publisher} {Springer},\ \bibinfo {year} {1992})\BibitemShut {NoStop}%
\bibitem [{\citenamefont {Fisher}(1967)}]{Fisher1967}%
  \BibitemOpen
  \bibfield  {author} {\bibinfo {author} {\bibfnamefont {M.~E.}\ \bibnamefont
  {Fisher}},\ }\bibfield  {title} {\bibinfo {title} {{The theory of equilibrium
  critical phenomena}},\ }\href {https://doi.org/10.1088/0034-4885/30/2/306}
  {\bibfield  {journal} {\bibinfo  {journal} {Rep. Prog. Phys.}\ }\textbf
  {\bibinfo {volume} {30}},\ \bibinfo {pages} {615} (\bibinfo {year}
  {1967})}\BibitemShut {NoStop}%
\bibitem [{\citenamefont {Agrawal}\ \emph {et~al.}(2021)\citenamefont
  {Agrawal}, \citenamefont {Kumar},\ and\ \citenamefont {Puri}}]{Puri2021}%
  \BibitemOpen
  \bibfield  {author} {\bibinfo {author} {\bibfnamefont {R.}~\bibnamefont
  {Agrawal}}, \bibinfo {author} {\bibfnamefont {M.}~\bibnamefont {Kumar}},\
  and\ \bibinfo {author} {\bibfnamefont {S.}~\bibnamefont {Puri}},\ }\bibfield
  {title} {\bibinfo {title} {{Domain growth and aging in the random field $XY$
  model: A Monte Carlo study}},\ }\href
  {https://doi.org/10.1103/PhysRevE.104.044123} {\bibfield  {journal} {\bibinfo
   {journal} {Phys. Rev. E}\ }\textbf {\bibinfo {volume} {104}},\ \bibinfo
  {pages} {044123} (\bibinfo {year} {2021})}\BibitemShut {NoStop}%
\bibitem [{\citenamefont {Hasenbusch}\ and\ \citenamefont
  {Meyer}(1990)}]{HASENBUSCH1990238}%
  \BibitemOpen
  \bibfield  {author} {\bibinfo {author} {\bibfnamefont {M.}~\bibnamefont
  {Hasenbusch}}\ and\ \bibinfo {author} {\bibfnamefont {S.}~\bibnamefont
  {Meyer}},\ }\bibfield  {title} {\bibinfo {title} {{Critical exponents of the
  3D XY model from cluster update Monte Carlo}},\ }\href
  {https://doi.org/https://doi.org/10.1016/0370-2693(90)91286-K} {\bibfield
  {journal} {\bibinfo  {journal} {Phys. Lett. B}\ }\textbf {\bibinfo {volume}
  {241}},\ \bibinfo {pages} {238} (\bibinfo {year} {1990})}\BibitemShut
  {NoStop}%
\bibitem [{\citenamefont {Gottlob}\ and\ \citenamefont
  {Hasenbusch}(1993)}]{Gottlob}%
  \BibitemOpen
  \bibfield  {author} {\bibinfo {author} {\bibfnamefont {A.~P.}\ \bibnamefont
  {Gottlob}}\ and\ \bibinfo {author} {\bibfnamefont {M.}~\bibnamefont
  {Hasenbusch}},\ }\bibfield  {title} {\bibinfo {title} {{{Critical behavior of
  the 3D XY model: A Monte Carlo study}}},\ }\href
  {https://doi.org/10.1016/0378-4371(93)90131-M} {\bibfield  {journal}
  {\bibinfo  {journal} {Physica A}\ }\textbf {\bibinfo {volume} {201}},\
  \bibinfo {pages} {593} (\bibinfo {year} {1993})}\BibitemShut {NoStop}%
\bibitem [{\citenamefont {Campostrini}\ \emph {et~al.}(2001)\citenamefont
  {Campostrini}, \citenamefont {Hasenbusch}, \citenamefont {Pelissetto},
  \citenamefont {Rossi},\ and\ \citenamefont {Vicari}}]{Hasenbusch2001}%
  \BibitemOpen
  \bibfield  {author} {\bibinfo {author} {\bibfnamefont {M.}~\bibnamefont
  {Campostrini}}, \bibinfo {author} {\bibfnamefont {M.}~\bibnamefont
  {Hasenbusch}}, \bibinfo {author} {\bibfnamefont {A.}~\bibnamefont
  {Pelissetto}}, \bibinfo {author} {\bibfnamefont {P.}~\bibnamefont {Rossi}},\
  and\ \bibinfo {author} {\bibfnamefont {E.}~\bibnamefont {Vicari}},\
  }\bibfield  {title} {\bibinfo {title} {{Critical behavior of the
  three-dimensional $\mathrm{XY}$ universality class}},\ }\href
  {https://doi.org/10.1103/PhysRevB.63.214503} {\bibfield  {journal} {\bibinfo
  {journal} {Phys. Rev. B}\ }\textbf {\bibinfo {volume} {63}},\ \bibinfo
  {pages} {214503} (\bibinfo {year} {2001})}\BibitemShut {NoStop}%
\bibitem [{\citenamefont {Bray}(2002)}]{Bray}%
  \BibitemOpen
  \bibfield  {author} {\bibinfo {author} {\bibfnamefont {A.~J.}\ \bibnamefont
  {Bray}},\ }\bibfield  {title} {\bibinfo {title} {{Theory of phase-ordering
  kinetics}},\ }\href
  {https://doi.org/https://doi.org/10.1080/00018730110117433} {\bibfield
  {journal} {\bibinfo  {journal} {Adv. Phys.}\ }\textbf {\bibinfo {volume}
  {51}},\ \bibinfo {pages} {481} (\bibinfo {year} {2002})}\BibitemShut
  {NoStop}%
\bibitem [{\citenamefont {Puri}\ and\ \citenamefont
  {Wadhawan}(2009)}]{PuriWadhawan2009}%
  \BibitemOpen
  \bibinfo {editor} {\bibfnamefont {S.}~\bibnamefont {Puri}}\ and\ \bibinfo
  {editor} {\bibfnamefont {V.}~\bibnamefont {Wadhawan}},\ eds.,\ \href@noop {}
  {\emph {\bibinfo {title} {{Kinetics of Phase Transitions}}}}\ (\bibinfo
  {publisher} {CRC Press},\ \bibinfo {address} {Boca Raton},\ \bibinfo {year}
  {2009})\BibitemShut {NoStop}%
\bibitem [{\citenamefont {Kohring}\ \emph {et~al.}(1986)\citenamefont
  {Kohring}, \citenamefont {Shrock},\ and\ \citenamefont
  {Wills}}]{Peter_Wills}%
  \BibitemOpen
  \bibfield  {author} {\bibinfo {author} {\bibfnamefont {G.}~\bibnamefont
  {Kohring}}, \bibinfo {author} {\bibfnamefont {R.~E.}\ \bibnamefont
  {Shrock}},\ and\ \bibinfo {author} {\bibfnamefont {P.}~\bibnamefont
  {Wills}},\ }\bibfield  {title} {\bibinfo {title} {{Role of Vortex Strings in
  the Three-Dimensional O(2) Model}},\ }\href
  {https://doi.org/10.1103/PhysRevLett.57.1358} {\bibfield  {journal} {\bibinfo
   {journal} {Phys. Rev. Lett.}\ }\textbf {\bibinfo {volume} {57}},\ \bibinfo
  {pages} {1358} (\bibinfo {year} {1986})}\BibitemShut {NoStop}%
\bibitem [{\citenamefont {Chaikin}\ \emph {et~al.}(1995)\citenamefont
  {Chaikin}, \citenamefont {Lubensky},\ and\ \citenamefont
  {Witten}}]{chaikin1995}%
  \BibitemOpen
  \bibfield  {author} {\bibinfo {author} {\bibfnamefont {P.~M.}\ \bibnamefont
  {Chaikin}}, \bibinfo {author} {\bibfnamefont {T.~C.}\ \bibnamefont
  {Lubensky}},\ and\ \bibinfo {author} {\bibfnamefont {T.~A.}\ \bibnamefont
  {Witten}},\ }\href@noop {} {\emph {\bibinfo {title} {{Principles of condensed
  matter physics}}}},\ Vol.~\bibinfo {volume} {10}\ (\bibinfo  {publisher}
  {Cambridge university press Cambridge},\ \bibinfo {year} {1995})\BibitemShut
  {NoStop}%
\bibitem [{\citenamefont {Ren}\ \emph {et~al.}(2007)\citenamefont {Ren},
  \citenamefont {O'Keeffe},\ and\ \citenamefont {Orkoulas}}]{Orkoulas2007}%
  \BibitemOpen
  \bibfield  {author} {\bibinfo {author} {\bibfnamefont {R.}~\bibnamefont
  {Ren}}, \bibinfo {author} {\bibfnamefont {C.~J.}\ \bibnamefont {O'Keeffe}},\
  and\ \bibinfo {author} {\bibfnamefont {G.}~\bibnamefont {Orkoulas}},\
  }\bibfield  {title} {\bibinfo {title} {Sequential updating algorithms for
  grand canonical monte carlo simulations},\ }\href
  {https://doi.org/10.1080/00268970601143341} {\bibfield  {journal} {\bibinfo
  {journal} {Mol. Phys.}\ }\textbf {\bibinfo {volume} {105}},\ \bibinfo {pages}
  {231} (\bibinfo {year} {2007})}\BibitemShut {NoStop}%
\bibitem [{\citenamefont {Kumar}\ \emph {et~al.}(2017)\citenamefont {Kumar},
  \citenamefont {Chatterjee}, \citenamefont {Paul},\ and\ \citenamefont
  {Puri}}]{Puri2017}%
  \BibitemOpen
  \bibfield  {author} {\bibinfo {author} {\bibfnamefont {M.}~\bibnamefont
  {Kumar}}, \bibinfo {author} {\bibfnamefont {S.}~\bibnamefont {Chatterjee}},
  \bibinfo {author} {\bibfnamefont {R.}~\bibnamefont {Paul}},\ and\ \bibinfo
  {author} {\bibfnamefont {S.}~\bibnamefont {Puri}},\ }\bibfield  {title}
  {\bibinfo {title} {{Ordering kinetics in the random-bond $XY$ model}},\
  }\href {https://doi.org/10.1103/PhysRevE.96.042127} {\bibfield  {journal}
  {\bibinfo  {journal} {Phys. Rev. E}\ }\textbf {\bibinfo {volume} {96}},\
  \bibinfo {pages} {042127} (\bibinfo {year} {2017})}\BibitemShut {NoStop}%
\bibitem [{\citenamefont {Glauber}(1963)}]{Glauber1963}%
  \BibitemOpen
  \bibfield  {author} {\bibinfo {author} {\bibfnamefont {R.~J.}\ \bibnamefont
  {Glauber}},\ }\bibfield  {title} {\bibinfo {title} {{Time-Dependent
  Statistics of the Ising Model}},\ }\href {https://doi.org/10.1063/1.1703954}
  {\bibfield  {journal} {\bibinfo  {journal} {J. Math. Phys.}\ }\textbf
  {\bibinfo {volume} {4}},\ \bibinfo {pages} {294} (\bibinfo {year}
  {1963})}\BibitemShut {NoStop}%
\bibitem [{\citenamefont {Bray}\ and\ \citenamefont {Puri}(1991)}]{BP}%
  \BibitemOpen
  \bibfield  {author} {\bibinfo {author} {\bibfnamefont {A.~J.}\ \bibnamefont
  {Bray}}\ and\ \bibinfo {author} {\bibfnamefont {S.}~\bibnamefont {Puri}},\
  }\bibfield  {title} {\bibinfo {title} {{Asymptotic structure factor and
  power-law tails for phase ordering in systems with continuous symmetry}},\
  }\href {https://doi.org/10.1103/PhysRevLett.67.2670} {\bibfield  {journal}
  {\bibinfo  {journal} {Phys. Rev. Lett.}\ }\textbf {\bibinfo {volume} {67}},\
  \bibinfo {pages} {2670} (\bibinfo {year} {1991})}\BibitemShut {NoStop}%
\bibitem [{\citenamefont {Toyoki}(1992)}]{Toyoki}%
  \BibitemOpen
  \bibfield  {author} {\bibinfo {author} {\bibfnamefont {H.}~\bibnamefont
  {Toyoki}},\ }\bibfield  {title} {\bibinfo {title} {{Structure factors of
  vector-order-parameter systems containing random topological defects}},\
  }\href {https://doi.org/10.1103/PhysRevB.45.1965} {\bibfield  {journal}
  {\bibinfo  {journal} {Phys. Rev. B}\ }\textbf {\bibinfo {volume} {45}},\
  \bibinfo {pages} {1965} (\bibinfo {year} {1992})}\BibitemShut {NoStop}%
\bibitem [{\citenamefont {Majumder}\ and\ \citenamefont
  {Das}(2011)}]{Majumdar}%
  \BibitemOpen
  \bibfield  {author} {\bibinfo {author} {\bibfnamefont {S.}~\bibnamefont
  {Majumder}}\ and\ \bibinfo {author} {\bibfnamefont {S.~K.}\ \bibnamefont
  {Das}},\ }\bibfield  {title} {\bibinfo {title} {{Diffusive domain coarsening:
  Early time dynamics and finite-size effects}},\ }\href
  {https://doi.org/10.1103/PhysRevE.84.021110} {\bibfield  {journal} {\bibinfo
  {journal} {Phys. Rev. E}\ }\textbf {\bibinfo {volume} {84}},\ \bibinfo
  {pages} {021110} (\bibinfo {year} {2011})}\BibitemShut {NoStop}%
\bibitem [{\citenamefont {Vadakkayil}\ \emph {et~al.}(2022)\citenamefont
  {Vadakkayil}, \citenamefont {Singha},\ and\ \citenamefont
  {Das}}]{roughening_NV}%
  \BibitemOpen
  \bibfield  {author} {\bibinfo {author} {\bibfnamefont {N.}~\bibnamefont
  {Vadakkayil}}, \bibinfo {author} {\bibfnamefont {S.~K.}\ \bibnamefont
  {Singha}},\ and\ \bibinfo {author} {\bibfnamefont {S.~K.}\ \bibnamefont
  {Das}},\ }\bibfield  {title} {\bibinfo {title} {{Influence of roughening
  transition on magnetic ordering}},\ }\href
  {https://doi.org/10.1103/PhysRevE.105.044142} {\bibfield  {journal} {\bibinfo
   {journal} {Phys. Rev. E}\ }\textbf {\bibinfo {volume} {105}},\ \bibinfo
  {pages} {044142} (\bibinfo {year} {2022})}\BibitemShut {NoStop}%
\bibitem [{\citenamefont {Das}(2013)}]{SKD_pre2013}%
  \BibitemOpen
  \bibfield  {author} {\bibinfo {author} {\bibfnamefont {S.~K.}\ \bibnamefont
  {Das}},\ }\bibfield  {title} {\bibinfo {title} {{Unlocking of frozen dynamics
  in the complex Ginzburg-Landau equation}},\ }\href
  {https://doi.org/10.1103/PhysRevE.87.012135} {\bibfield  {journal} {\bibinfo
  {journal} {Phys. Rev. E}\ }\textbf {\bibinfo {volume} {87}},\ \bibinfo
  {pages} {012135} (\bibinfo {year} {2013})}\BibitemShut {NoStop}%
\bibitem [{\citenamefont {Mondello}\ and\ \citenamefont
  {Goldenfeld}(1990)}]{Goldenfeld_2d}%
  \BibitemOpen
  \bibfield  {author} {\bibinfo {author} {\bibfnamefont {M.}~\bibnamefont
  {Mondello}}\ and\ \bibinfo {author} {\bibfnamefont {N.}~\bibnamefont
  {Goldenfeld}},\ }\bibfield  {title} {\bibinfo {title} {{Scaling and vortex
  dynamics after the quench of a system with a continuous symmetry}},\ }\href
  {https://doi.org/10.1103/PhysRevA.42.5865} {\bibfield  {journal} {\bibinfo
  {journal} {Phys. Rev. A}\ }\textbf {\bibinfo {volume} {42}},\ \bibinfo
  {pages} {5865} (\bibinfo {year} {1990})}\BibitemShut {NoStop}%
\bibitem [{\citenamefont {Mondello}\ and\ \citenamefont
  {Goldenfeld}(1992)}]{Goldenfeld_3d}%
  \BibitemOpen
  \bibfield  {author} {\bibinfo {author} {\bibfnamefont {M.}~\bibnamefont
  {Mondello}}\ and\ \bibinfo {author} {\bibfnamefont {N.}~\bibnamefont
  {Goldenfeld}},\ }\bibfield  {title} {\bibinfo {title} {{Scaling and
  vortex-string dynamics in a three-dimensional system with a continuous
  symmetry}},\ }\href {https://doi.org/10.1103/PhysRevA.45.657} {\bibfield
  {journal} {\bibinfo  {journal} {Phys. Rev. A}\ }\textbf {\bibinfo {volume}
  {45}},\ \bibinfo {pages} {657} (\bibinfo {year} {1992})}\BibitemShut
  {NoStop}%
\bibitem [{\citenamefont {T{\"a}uber}(2017)}]{Tauber2017}%
  \BibitemOpen
  \bibfield  {author} {\bibinfo {author} {\bibfnamefont {U.~C.}\ \bibnamefont
  {T{\"a}uber}},\ }\bibfield  {title} {\bibinfo {title} {Phase transitions and
  scaling in systems far from equilibrium},\ }\href
  {https://doi.org/10.1146/annurev-conmatphys-031016-025444} {\bibfield
  {journal} {\bibinfo  {journal} {Annu. Rev. Condens. Matter Phys.}\ }\textbf
  {\bibinfo {volume} {8}},\ \bibinfo {pages} {185} (\bibinfo {year}
  {2017})}\BibitemShut {NoStop}%
\bibitem [{\citenamefont {Bray}\ and\ \citenamefont
  {Rutenberg}(1994)}]{Bray_Rutenberg}%
  \BibitemOpen
  \bibfield  {author} {\bibinfo {author} {\bibfnamefont {A.~J.}\ \bibnamefont
  {Bray}}\ and\ \bibinfo {author} {\bibfnamefont {A.~D.}\ \bibnamefont
  {Rutenberg}},\ }\bibfield  {title} {\bibinfo {title} {{Growth laws for phase
  ordering}},\ }\href {https://doi.org/10.1103/PhysRevE.49.R27} {\bibfield
  {journal} {\bibinfo  {journal} {Phys. Rev. E}\ }\textbf {\bibinfo {volume}
  {49}},\ \bibinfo {pages} {R27} (\bibinfo {year} {1994})}\BibitemShut
  {NoStop}%
\bibitem [{\citenamefont {Blundell}\ and\ \citenamefont
  {Bray}(1994)}]{Blundell_Bray}%
  \BibitemOpen
  \bibfield  {author} {\bibinfo {author} {\bibfnamefont {R.~E.}\ \bibnamefont
  {Blundell}}\ and\ \bibinfo {author} {\bibfnamefont {A.~J.}\ \bibnamefont
  {Bray}},\ }\bibfield  {title} {\bibinfo {title} {{Phase-ordering dynamics of
  the O(n) model: Exact predictions and numerical results}},\ }\href
  {https://doi.org/10.1103/PhysRevE.49.4925} {\bibfield  {journal} {\bibinfo
  {journal} {Phys. Rev. E}\ }\textbf {\bibinfo {volume} {49}},\ \bibinfo
  {pages} {4925} (\bibinfo {year} {1994})}\BibitemShut {NoStop}%
\bibitem [{\citenamefont {Huse}(1986)}]{Huse}%
  \BibitemOpen
  \bibfield  {author} {\bibinfo {author} {\bibfnamefont {D.~A.}\ \bibnamefont
  {Huse}},\ }\bibfield  {title} {\bibinfo {title} {{Corrections to late-stage
  behavior in spinodal decomposition: Lifshitz-Slyozov scaling and Monte Carlo
  simulations}},\ }\href {https://doi.org/10.1103/PhysRevB.34.7845} {\bibfield
  {journal} {\bibinfo  {journal} {Phys. Rev. B}\ }\textbf {\bibinfo {volume}
  {34}},\ \bibinfo {pages} {7845} (\bibinfo {year} {1986})}\BibitemShut
  {NoStop}%
\bibitem [{\citenamefont {Olejarz}\ \emph {et~al.}(2011)\citenamefont
  {Olejarz}, \citenamefont {Krapivsky},\ and\ \citenamefont
  {Redner}}]{zerotmp_redner}%
  \BibitemOpen
  \bibfield  {author} {\bibinfo {author} {\bibfnamefont {J.}~\bibnamefont
  {Olejarz}}, \bibinfo {author} {\bibfnamefont {P.~L.}\ \bibnamefont
  {Krapivsky}},\ and\ \bibinfo {author} {\bibfnamefont {S.}~\bibnamefont
  {Redner}},\ }\bibfield  {title} {\bibinfo {title} {{Zero-temperature
  relaxation of three-dimensional Ising ferromagnets}},\ }\href
  {https://doi.org/10.1103/PhysRevE.83.051104} {\bibfield  {journal} {\bibinfo
  {journal} {Phys. Rev. E}\ }\textbf {\bibinfo {volume} {83}},\ \bibinfo
  {pages} {051104} (\bibinfo {year} {2011})}\BibitemShut {NoStop}%
\bibitem [{\citenamefont {Vadakkayil}\ \emph {et~al.}(2019)\citenamefont
  {Vadakkayil}, \citenamefont {Chakraborty},\ and\ \citenamefont
  {Das}}]{zerotmp_NV}%
  \BibitemOpen
  \bibfield  {author} {\bibinfo {author} {\bibfnamefont {N.}~\bibnamefont
  {Vadakkayil}}, \bibinfo {author} {\bibfnamefont {S.}~\bibnamefont
  {Chakraborty}},\ and\ \bibinfo {author} {\bibfnamefont {S.~K.}\ \bibnamefont
  {Das}},\ }\bibfield  {title} {\bibinfo {title} {{Finite-size scaling study of
  aging during coarsening in non-conserved Ising model: The case of zero
  temperature quench}},\ }\href {https://doi.org/10.1063/1.5052418} {\bibfield
  {journal} {\bibinfo  {journal} {J. Chem. Phys.}\ }\textbf {\bibinfo {volume}
  {150}},\ \bibinfo {pages} {054702} (\bibinfo {year} {2019})}\BibitemShut
  {NoStop}%
\bibitem [{\citenamefont {Gessert}\ \emph {et~al.}(2024)\citenamefont
  {Gessert}, \citenamefont {Christiansen},\ and\ \citenamefont
  {Janke}}]{Janke_zero}%
  \BibitemOpen
  \bibfield  {author} {\bibinfo {author} {\bibfnamefont {D.}~\bibnamefont
  {Gessert}}, \bibinfo {author} {\bibfnamefont {H.}~\bibnamefont
  {Christiansen}},\ and\ \bibinfo {author} {\bibfnamefont {W.}~\bibnamefont
  {Janke}},\ }\bibfield  {title} {\bibinfo {title} {{Aging following a
  zero-temperature quench in the $d=3$ Ising model}},\ }\href
  {https://doi.org/10.1103/PhysRevE.109.044148} {\bibfield  {journal} {\bibinfo
   {journal} {Phys. Rev. E}\ }\textbf {\bibinfo {volume} {109}},\ \bibinfo
  {pages} {044148} (\bibinfo {year} {2024})}\BibitemShut {NoStop}%
\bibitem [{\citenamefont {Chakraborty}\ and\ \citenamefont
  {Das}(2017)}]{Chakraborty2017}%
  \BibitemOpen
  \bibfield  {author} {\bibinfo {author} {\bibfnamefont {S.}~\bibnamefont
  {Chakraborty}}\ and\ \bibinfo {author} {\bibfnamefont {S.~K.}\ \bibnamefont
  {Das}},\ }\bibfield  {title} {\bibinfo {title} {Coarsening in {3D}
  nonconserved {Ising} model at zero temperature: Anomaly in structure and slow
  relaxation of order-parameter autocorrelation},\ }\href
  {https://doi.org/10.1209/0295-5075/119/50005} {\bibfield  {journal} {\bibinfo
   {journal} {Europhys. Lett.}\ }\textbf {\bibinfo {volume} {119}},\ \bibinfo
  {pages} {50005} (\bibinfo {year} {2017})}\BibitemShut {NoStop}%
\bibitem [{\citenamefont {Spirin}\ \emph {et~al.}(2001)\citenamefont {Spirin},
  \citenamefont {Krapivsky},\ and\ \citenamefont {Redner}}]{Redner2001}%
  \BibitemOpen
  \bibfield  {author} {\bibinfo {author} {\bibfnamefont {V.}~\bibnamefont
  {Spirin}}, \bibinfo {author} {\bibfnamefont {P.}~\bibnamefont {Krapivsky}},\
  and\ \bibinfo {author} {\bibfnamefont {S.}~\bibnamefont {Redner}},\
  }\bibfield  {title} {\bibinfo {title} {{Freezing in Ising ferromagnets}},\
  }\href {https://doi.org/10.1103/PhysRevE.65.016119} {\bibfield  {journal}
  {\bibinfo  {journal} {Phys. Rev. E}\ }\textbf {\bibinfo {volume} {65}},\
  \bibinfo {pages} {016119} (\bibinfo {year} {2001})}\BibitemShut {NoStop}%
\bibitem [{\citenamefont {Das}\ and\ \citenamefont
  {Chakraborty}(2017)}]{Das2017}%
  \BibitemOpen
  \bibfield  {author} {\bibinfo {author} {\bibfnamefont {S.~K.}\ \bibnamefont
  {Das}}\ and\ \bibinfo {author} {\bibfnamefont {S.}~\bibnamefont
  {Chakraborty}},\ }\bibfield  {title} {\bibinfo {title} {{Kinetics of
  ferromagnetic ordering in 3D Ising model: How far do we understand the case
  of a zero temperature quench?}},\ }\href
  {https://doi.org/10.1140/epjst/e2016-60313-6} {\bibfield  {journal} {\bibinfo
   {journal} {Eur. Phys. J. Spec. Top.}\ }\textbf {\bibinfo {volume} {226}},\
  \bibinfo {pages} {765} (\bibinfo {year} {2017})}\BibitemShut {NoStop}%
\bibitem [{\citenamefont {Hohenberg}\ and\ \citenamefont
  {Halperin}(1977)}]{Hohenberg}%
  \BibitemOpen
  \bibfield  {author} {\bibinfo {author} {\bibfnamefont {P.~C.}\ \bibnamefont
  {Hohenberg}}\ and\ \bibinfo {author} {\bibfnamefont {B.~I.}\ \bibnamefont
  {Halperin}},\ }\bibfield  {title} {\bibinfo {title} {{Theory of dynamic
  critical phenomena}},\ }\href {https://doi.org/10.1103/RevModPhys.49.435}
  {\bibfield  {journal} {\bibinfo  {journal} {Rev. Mod. Phys.}\ }\textbf
  {\bibinfo {volume} {49}},\ \bibinfo {pages} {435} (\bibinfo {year}
  {1977})}\BibitemShut {NoStop}%
\bibitem [{\citenamefont {Stanley}(1987)}]{Stanley}%
  \BibitemOpen
  \bibfield  {author} {\bibinfo {author} {\bibfnamefont {H.~E.}\ \bibnamefont
  {Stanley}},\ }\href@noop {} {\emph {\bibinfo {title} {Introduction to Phase
  Transitions and Critical Phenomena}}}\ (\bibinfo  {publisher} {Oxford
  University Press},\ \bibinfo {address} {Oxford},\ \bibinfo {year}
  {1987})\BibitemShut {NoStop}%
\bibitem [{\citenamefont {Onuki}(2002)}]{Onuki2002}%
  \BibitemOpen
  \bibfield  {author} {\bibinfo {author} {\bibfnamefont {A.}~\bibnamefont
  {Onuki}},\ }\href@noop {} {\emph {\bibinfo {title} {Phase Transition
  Dynamics}}}\ (\bibinfo  {publisher} {Cambridge University Press},\ \bibinfo
  {year} {2002})\BibitemShut {NoStop}%
\bibitem [{\citenamefont {Fisher}(1964)}]{Fisher1964}%
  \BibitemOpen
  \bibfield  {author} {\bibinfo {author} {\bibfnamefont {M.~E.}\ \bibnamefont
  {Fisher}},\ }\bibfield  {title} {\bibinfo {title} {{Correlation Functions and
  the Critical Region of Simple Fluids}},\ }\href
  {https://doi.org/10.1063/1.1704197} {\bibfield  {journal} {\bibinfo
  {journal} {J. Math. Phys.}\ }\textbf {\bibinfo {volume} {5}},\ \bibinfo
  {pages} {944} (\bibinfo {year} {1964})}\BibitemShut {NoStop}%
\bibitem [{\citenamefont {Kadanoff}(1966)}]{Kadanoff}%
  \BibitemOpen
  \bibfield  {author} {\bibinfo {author} {\bibfnamefont {L.~P.}\ \bibnamefont
  {Kadanoff}},\ }\bibfield  {title} {\bibinfo {title} {{Scaling laws for Ising
  models near ${T}_{c}$}},\ }\href
  {https://doi.org/10.1103/PhysicsPhysiqueFizika.2.263} {\bibfield  {journal}
  {\bibinfo  {journal} {Phys. Phys. Fiz.}\ }\textbf {\bibinfo {volume} {2}},\
  \bibinfo {pages} {263} (\bibinfo {year} {1966})}\BibitemShut {NoStop}%
\bibitem [{\citenamefont {Zinn-Justin}(2001)}]{ZinnJustin2001}%
  \BibitemOpen
  \bibfield  {author} {\bibinfo {author} {\bibfnamefont {J.}~\bibnamefont
  {Zinn-Justin}},\ }\bibfield  {title} {\bibinfo {title} {{Precise
  determination of critical exponents and equation of state by field theory
  methods}},\ }\href {https://doi.org/10.1016/S0370-1573(00)00126-5} {\bibfield
   {journal} {\bibinfo  {journal} {Phys. Rep.}\ }\textbf {\bibinfo {volume}
  {344}},\ \bibinfo {pages} {159} (\bibinfo {year} {2001})}\BibitemShut
  {NoStop}%
\bibitem [{\citenamefont {Efron}(1982)}]{efron_jackknife}%
  \BibitemOpen
  \bibfield  {author} {\bibinfo {author} {\bibfnamefont {B.}~\bibnamefont
  {Efron}},\ }\href@noop {} {\emph {\bibinfo {title} {{The jackknife, the
  bootstrap and other resampling plans}}}}\ (\bibinfo  {publisher} {SIAM},\
  \bibinfo {year} {1982})\BibitemShut {NoStop}%
\bibitem [{\citenamefont {Kosterlitz}\ and\ \citenamefont
  {Thouless}(1973)}]{Kosterlitz1973}%
  \BibitemOpen
  \bibfield  {author} {\bibinfo {author} {\bibfnamefont {J.~M.}\ \bibnamefont
  {Kosterlitz}}\ and\ \bibinfo {author} {\bibfnamefont {D.~J.}\ \bibnamefont
  {Thouless}},\ }\bibfield  {title} {\bibinfo {title} {{Ordering, metastability
  and phase transitions in two-dimensional systems}},\ }\href
  {https://doi.org/10.1088/0022-3719/6/7/010} {\bibfield  {journal} {\bibinfo
  {journal} {J. Phys. C}\ }\textbf {\bibinfo {volume} {6}},\ \bibinfo {pages}
  {1181} (\bibinfo {year} {1973})}\BibitemShut {NoStop}%
\bibitem [{\citenamefont {Kosterlitz}(1974)}]{Kosterlitz1974}%
  \BibitemOpen
  \bibfield  {author} {\bibinfo {author} {\bibfnamefont {J.~M.}\ \bibnamefont
  {Kosterlitz}},\ }\bibfield  {title} {\bibinfo {title} {{The critical
  properties of the two-dimensional {XY} model}},\ }\href
  {https://doi.org/10.1088/0022-3719/7/6/005} {\bibfield  {journal} {\bibinfo
  {journal} {J. Phys. C}\ }\textbf {\bibinfo {volume} {7}},\ \bibinfo {pages}
  {1046} (\bibinfo {year} {1974})}\BibitemShut {NoStop}%
\end{thebibliography}
\end{document}